\documentclass{aastex62}

\submitjournal{ApJ}

\shorttitle{HO\,Puppis}
\shortauthors{Lee et al.}

\begin{document}

\title{HO\,Puppis: Not a Be Star but a Newly Confirmed IW And-Type Star}

\correspondingauthor{Chow-Choong Ngeow}
\email{cngeow@astro.ncu.edu.tw}

\author[0000-0002-3142-7299]{Chien-De Lee}
\author{Jia-Yu Ou}
\affil{Graduate Institute of Astronomy, National Central University, Jhongli 32001, Taiwan}
\author[0000-0001-8894-0854]{Po-Chieh Yu}
\affil{Graduate Institute of Astronomy, National Central University, Jhongli 32001, Taiwan}
\affil{College of General Studies, Yuan-Ze University, Chung-Li 32003, Taiwan}
\author[0000-0001-8771-7554]{Chow-Choong Ngeow}
\affil{Graduate Institute of Astronomy, National Central University, Jhongli 32001, Taiwan}
\author{Po-Chieh Huang}
\affil{Graduate Institute of Astronomy, National Central University, Jhongli 32001, Taiwan}
\author{Wing-Huen Ip}
\affil{Graduate Institute of Astronomy, National Central University, Jhongli 32001, Taiwan}
\author{Franz-Josef Hambsch}
\affil{American Association of Variable Star Observers (AAVSO), Cambridge, MA, USA}
\affil{Vereniging Voor Sterrenkunde (VVS), Oostmeers 122 C, 8000 Brugge, Belgium}
\affil{Bundesdeutsche Arbeitsgemeinschaft für Veränderliche Sterne e.V. (BAV), Munsterdamm 90, D-12169 Berlin, Germany}
\author{Hyun-il Sung}
\affil{Korea Astronomy and Space Science Institute (KASI), Bohyunsan Optical Astronomy Observatory (BOAO), Youngcheon, Gyungbuk 38812,
Republic of Korea}
\author{Jan van Roestel}
\affiliation{Division of Physics, Mathematics, and Astronomy, California Institute of Technology, Pasadena, CA 91125, USA}

\author{Richard Dekany}
\affiliation{Caltech Optical Observatories, California Institute of Technology, Pasadena, CA 91125, USA}

\author{Andrew J. Drake}
\affiliation{Division of Physics, Mathematics, and Astronomy, California Institute of Technology, Pasadena, CA 91125, USA}
\author[0000-0002-3168-0139]{Matthew J. Graham}
\affiliation{Division of Physics, Mathematics, and Astronomy, California Institute of Technology, Pasadena, CA 91125, USA}

\author[0000-0001-5060-8733]{Dmitry A. Duev}
\affiliation{Division of Physics, Mathematics, and Astronomy, California Institute of Technology, Pasadena, CA 91125, USA}

\author{Stephen Kaye}
\affiliation{Caltech Optical Observatories, California Institute of Technology, Pasadena, CA 91125, USA}
\author[0000-0002-6540-1484]{Thomas Kupfer}
\affiliation{Kavli Institute for Theoretical Physics, University of California, Santa Barbara, CA 93106, USA}

\author[0000-0003-2451-5482]{Russ R. Laher}
\affiliation{IPAC, California Institute of Technology, Pasadena, CA 91125, USA}

\author[0000-0002-8532-9395]{Frank J. Masci}
\affiliation{IPAC, California Institute of Technology, Pasadena, CA 91125, USA}

\author[0000-0001-7016-1692]{Przemek Mr\'oz}
\affil{Division of Physics, Mathematics, and Astronomy, California Institute of Technology, Pasadena, CA 91125, USA}

\author[0000-0002-0466-1119]{James D. Neill}
\affiliation{Caltech Optical Observatories, California Institute of Technology, Pasadena, CA  91125, USA}

\author[0000-0002-0387-370X]{Reed Riddle}
\affiliation{Caltech Optical Observatories, California Institute of Technology, Pasadena, CA  91125, USA}

\author[0000-0001-7648-4142]{Ben Rusholme}
\affiliation{IPAC, California Institute of Technology, Pasadena, CA 91125, USA}

\author[0000-0002-1835-6078]{Richard Walters}
\affiliation{Caltech Optical Observatories, California Institute of Technology, Pasadena, CA 91125, USA}

\begin{abstract}
HO Puppis (HO Pup) was considered as a Be-star candidate based on its $\gamma$ Cassiopeiae-type light curve, but lacked spectroscopic confirmation. Using distance measured from Gaia Data Release 2 and the spectral-energy-distribution (SED) fit on broadband photometry, the Be-star nature of HO Pup is ruled out. Furthermore, based on the 28,700 photometric data points collected from various time-domain surveys and dedicated intensive-monitoring observations, the light curves of HO Pup closely resemble IW\,And-type stars \citep[as pointed out in][]{kim20}, exhibiting characteristics such as quasi-standstill phase, brightening, and dips. The light curve of HO Pup displays various variability timescales, including brightening cycles ranging from 23 to 61 days, variations with periods between 3.9 days and 50 minutes during the quasi-standstill phase, and a semi-regular $\sim14$-day period for the dip events. We have also collected time-series spectra (with various spectral resolutions), in which Balmer emission lines and other expected spectral lines for an IW\,And-type star were detected (even though some of these lines were also expected to be present for Be stars). We detect Bowen fluorescence near the brightening phase, and that can be used to discriminate between IW\,And-type stars and Be stars. Finally, despite only observing for four nights, the polarization variation was detected, indicating that HO Pup has significant intrinsic polarization.

\end{abstract}

\section{Introduction} \label{sec:intro}

Be phenomena are the photometric and spectroscopic variability seen in the main-sequence luminous rapid rotators, known as Be stars, with a luminosity class III$-$V. In recent years, we have studied the evolutionary effect on the formation of Be stars in open clusters \citep{yu15,yu16,yu18} using the Palomar Transient Factory \citep[PTF;][]{Law2009} and the intermediate Palomar Transient Factory \citep[iPTF;][]{kul13}. The Zwicky Transient Facility \citep[ZTF;][]{bel19,gra19,mas19} came after iPTF, and its improved data can extend our investigation on the variability of Be stars \citep{nge19}, especially for the Be stars and Be-star candidates at the faint end ($m>13$~ mag), which were largely excluded in previous works \citep[e.g., in][]{lab17}.  Together with accompanying time-series spectroscopic data, we have a new opportunity to explore the fundamental time-domain nature of Be stars.  

\begin{figure}
 \centering
 \includegraphics[width=1\columnwidth]{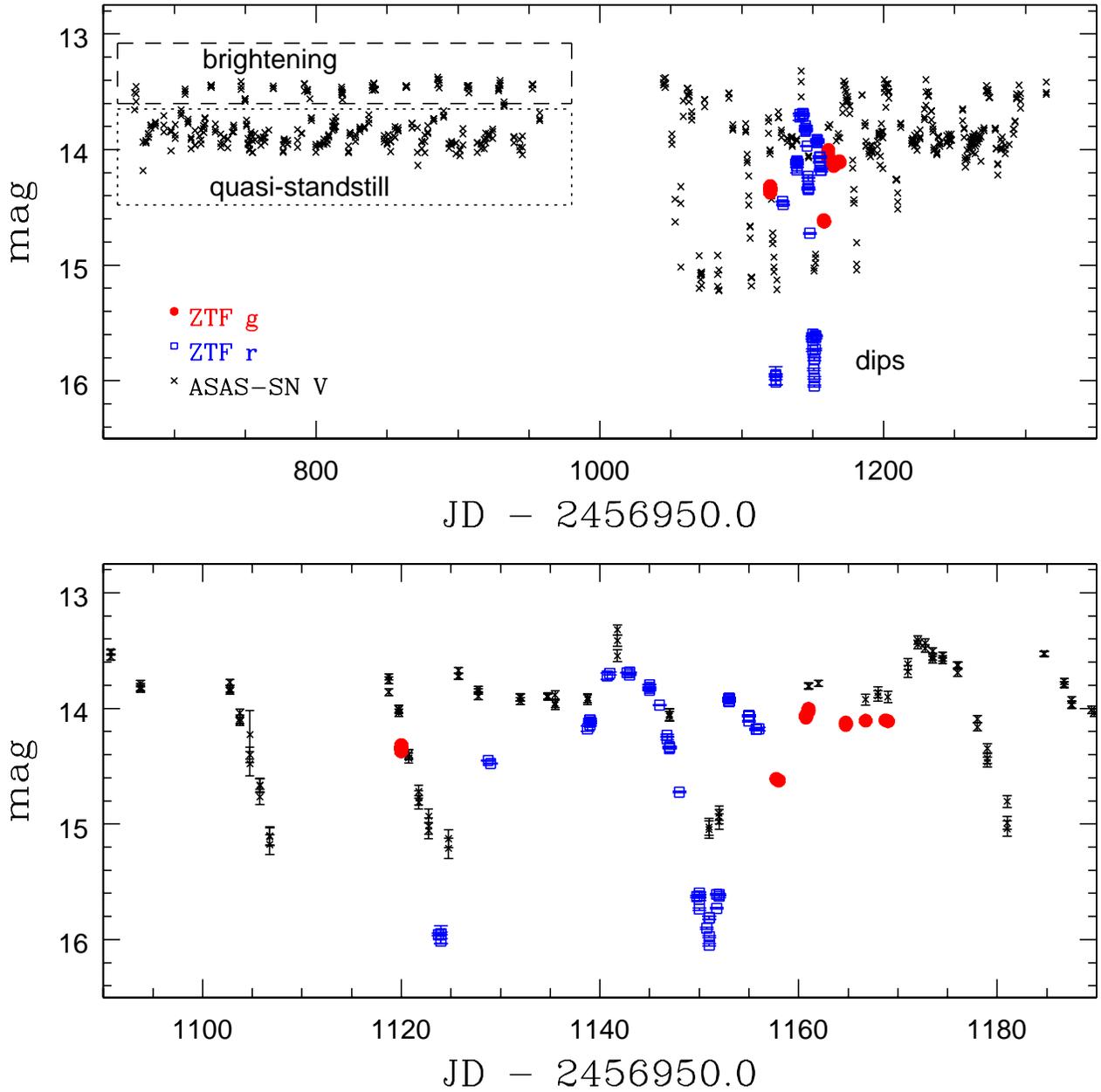}
	\caption{The ZTF g-band (red), r-band (blue) and ASAS-SN V-band (black) light curves of HO\,Pup.  The top panel shows a portion of the light curve across $\sim600$~days, together with terminologies that are used in this work to describe the light curve features: quasi-standstills (data points in the dotted box) are part of the light curves with magnitudes close to the mean value; brightenings are data points (in the dashed box) showing a brightening of $\sim0.5$~mag; and dips are events that fade by more than $\sim 1$~mag from the mean magnitude.  Several fading events with $\sim2.5$~mag dips were observed recently.  This epoch with a number of deep-dip events is enlarged in the bottom panel.  Consistency of the light curves seen in both ZTF and ASAS-SN observations ruled out the possibility that such dips were due to instrumental effects.   
	}
 \label{hopupztf}
\end{figure}

Here we report the photometric characteristics of HO\,Puppis (HO\,Pup,  $\alpha_{J2000}=7^{h}33^{m}54\fs18$,  $\delta_{J2000}=-15\degr45\arcmin38\farcs28$) as a result of our investigations of Be-stars variability with ZTF. HO\,Pup is listed as a Be star in the {\tt SIMBAD} database \citep{man97} and, hence, is included in our list of Be-star candidates, in which the classification is based on its $\gamma$ Cassiopeia (GCAS) type variability recorded by \citet[][]{sam17} -- a class of variable stars that exhibit eruptive irregular variability that is not easily classified further. Some early literature even suggested that HO\,Pup was a possible type Ia supernova with V-band photometry varying between 12.7 mag and 14.2 mag \citep{kuk71}. As presented in Figure~\ref{hopupztf}, two highly unusual $\sim2.5$-magnitude dips of HO\,Pup were observed by ZTF in 2017 November. In r-band, the ZTF data cover the full brightness minimum of the two dips at 16 mag from near the minimum to maximum brightness within two days because of the high-cadence sampling. These two events were also witnessed by the All-Sky Automated Survey for Supernovae \citep[ASAS-SN,][]{sha14,koc17} survey.  In addition to these two events recorded by ZTF and ASAS-SN, we have found more dips based on the literature and archival data since 2009 (see Section~\ref{sec:obs}). These dips with amplitudes of $\sim2.5$~mag are very unusual to be observed in typical Be stars, and hence HO\,Pup caught our attention and merits further investigation.

The variability behavior of HO\,Pup was also considered as a special type of cataclysmic variable (CV).  The ASAS-SN light curve of HO\,Pup was discussed in a recent work by \citet{kim20}.  In their study, HO\,Pup was classified as a dwarf nova (DN) with a unique heartbeat-like oscillation in its photometric variation, known as IW\,And-type stars. The stars that display IW\,And-type phenomenon represent a new sub-classification of Z\,Cam stars \citep{kat19}. The characteristics of IW\,And-type stars include eclipsing-like transients that happen once in a while, in addition to repeated brightening events seen in their light curves. Similar kinds of dip events followed by the outbursts or brightenings were also seen in several CVs.  Unlike the deep dips of HO\,Pup losing 90$\%$ of its starlight, other CVs tend to reduce only about 50$\%$ of its outburst brightness \citep{mas16,sch19}. At most, a reduction of 80$\%$ of the outburst brightness can be seen in the case of KIC\,9406652 \citep{gie13}. Besides the prototype DN, IW And, in the literature several DNe such as IM Eri, FY Vul, V507 Cyg, ST Cha, V513 Cas, and KIC 9406652 were classified as IW\,And-type stars, and the recent development of optical monitoring advances the identification \citep[see][and reference therein]{kim20,kim20b}.\footnote{More candidates for IW\,And-type stars can be found in VSNET Collaboration (http://ooruri.kusastro.kyoto-u.ac.jp/pipermail/vsnet-chat/). For example, V526\,Ori (vsnet-chat 8474), MGAB-V1252 (vsnet-chat 8101), USNO-A2.0 1275-09782989 (vsnet-chat 8432), RX J1831.7+6511 (vsnet-chat 24230), V2837 Ori (vsnet-chat 23538), LN\,UMa (vsnet-chat 24347), and EZ\,Vul (vsnet-chat 8266). } HO\,Pup could be an additional member of the IW\,And-type stars \citep[][see also vsnet-chat 8162 from VSNET Collaboration]{kim20}.

Given the ambiguous nature of HO\,Pup (either as a Be-star or a IW\,And-type star), we collected its light-curve data as much as possible from archival catalogs, some ongoing surveys, and new dedicated observations. These collections of light curves are presented in Section~\ref{sec:obs}. In addition to light-curve data, we have also made spectroscopic and polarimetric observations on HO\,Pup, described further in the same section. Analysis and results based on the observations are presented in Section~\ref{sec:r&a}, in which we also present the first emission-line spectra of HO\,Pup --  confirming its emission-line nature. In Section~\ref{sec:dis}, we discuss the scenarios for the observed unusual light-curve behaviors of HO\,Pup, and we summarize our findings in Section~\ref{sec:sum}.

\section{Observations and Data} \label{sec:obs}

\begin{figure}
 \centering
 \includegraphics[width=1\columnwidth]{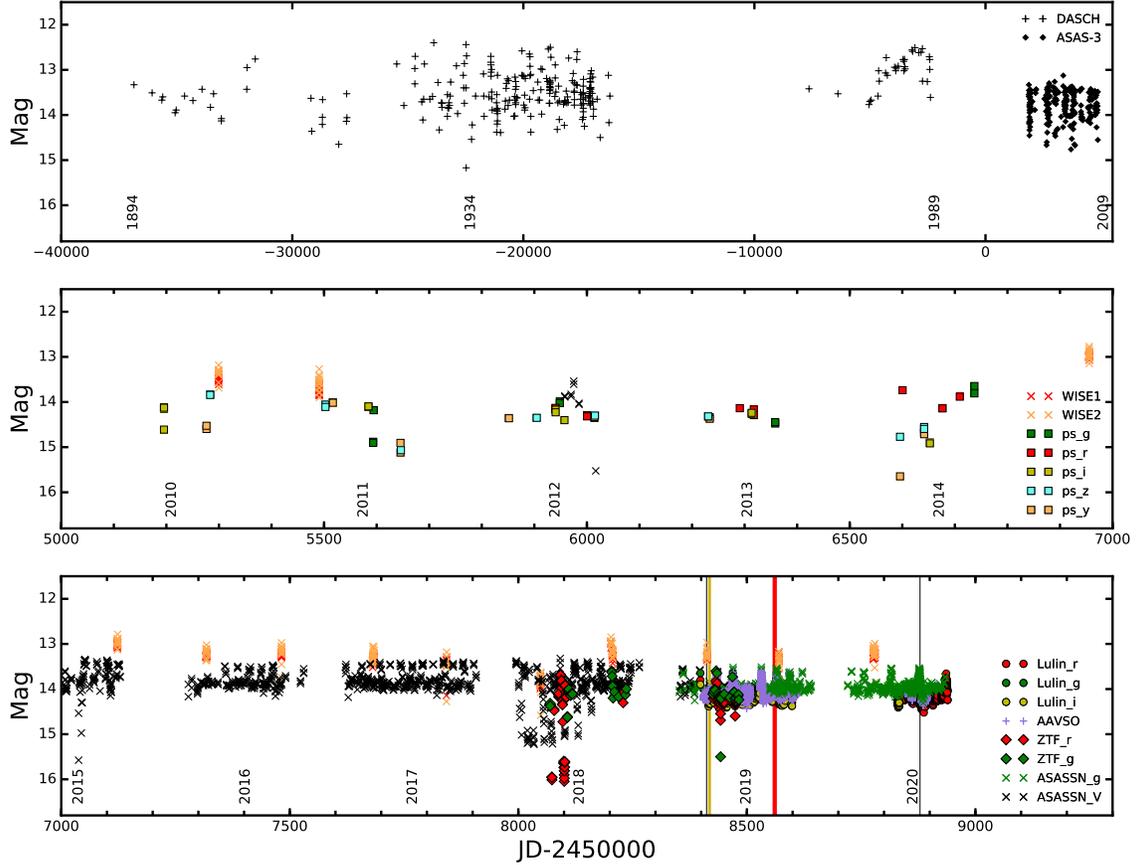}
	\caption{The light curve of HO\,Pup across more than a century from optical to mid-infrared, consisting of data from AAVSO, ASAS-3, ASAS-SN, DASCH, Lulin (with SLT), Pan-STARRS (PS), WISE and ZTF. Each data point (for clarity, error bars are ignored) is noted with data source and filters accordingly.  The observation time of follow-up spectroscopic and polarimetric observations are marked with vertical lines containing LOT/TRIPOL2 (yellow line in 2018), BOAO/BOES (blue line in 2018), CFHT/ESPaDOnS (red line in 2019) and P200/DBSP (black line in 2020).  More details can be seen in Figure~\ref{smallarea} with a focus on the individual duration.
	}
 \label{allLC}
\end{figure}

Time-series photometric data ranging from optical to infrared for HO\,Pup were collected from various survey catalogs, including the ZTF, the ASAS-SN, the Digital Access to a Sky Century @ Harvard \citep[DASCH,][]{gri12}, the third phase of the All Sky Automated Survey \citep[ASAS-3,][]{poj02}, the Panoramic Survey Telescope and Rapid Response System 3$\pi$ survey \citep[Pan-STARRS,][]{kai10,cha16}, the Wide-field Infrared Survey Explorer \citep[WISE,][]{cut12}, and observations available via the American Association of Variable Star Observers (AAVSO). These light-curve data were supplemented with dedicated observations taken at the Lulin Observatory in Taiwan.  All of these light curves were merged in Figure~\ref{allLC} and listed in Table~\ref{photlist}, which cover years from 1894 to 2020. We also summarized the light curves data in Table~\ref{phot}. Following the terminologies used to describe the light curves of IW\,And-type stars \citep{kim20}, definitions of main features exhibited in the light curves, such as dips, brightenings and quasi-standstills, are demonstrated in the upper panel of Figure~\ref{hopupztf}. Spectroscopic observations with low and high spectral resolutions were conducted by P60/SEDM, BOAO/BOES, CFHT/ESPaDOnS, and P200/DBSP. Additionally, polarization was measured in four different nights in late-October 2018 using the TRIPOL2 instrument installed at the Lulin Observatory. 

\begin{deluxetable}{l CCC CC}
\tabletypesize{\footnotesize}
\tablecaption{Collected light curves with 28700 data points for HO\,Pup. 
 \label{photlist}
			}
\tablecolumns{7}
\tablewidth{0pt}
\tablehead{ 
	\colhead{MJD} & \colhead{Mag} & \colhead{Uncertainty} & \colhead{Band} & \colhead{Source}
          }
\startdata
13146.3190&13.330 &0.100&B&   DASCH\\
13929.2218&13.510 &0.170&B&   DASCH\\
14353.0697&13.670 &0.170&B&   DASCH\\
14393.0297&13.600 &0.170&B&   DASCH\\
14927.3884&13.950 &0.120&B&   DASCH\\
14972.2725&13.890 &0.160&B&   DASCH\\
15336.3097&13.580 &0.150&B&   DASCH\\
15697.3473&13.680 &0.150&B&        DASCH\\
16093.2480&13.430 &0.090&B&        DASCH\\
16458.2944&13.830 &0.180&B&        DASCH \\
$\cdots$ & $\cdots$ & $\cdots$ & $\cdots$ & $\cdots$
\enddata
\tablecomments{The entire Table will be available in its electronic form at the {\tt SIMBAD} archive.}
\end{deluxetable}

\begin{deluxetable}{lccccccccc}
\tabletypesize{\footnotesize}
\tablecaption{Summary of light-curve data.   \label{phot}
			}
\tablecolumns{9}
\tablewidth{0pt}
\tablehead{ 
	\colhead{Band} & \colhead{Database} & \colhead{N} & \colhead{MJD (start)} & \colhead{MJD (end)} & \colhead{UT Date (start)} & \colhead{UT Date (end)} & \colhead{Mean Mag.} & \colhead{Std.}
          }
\startdata
B & DASCH      & 260 & 13146.3
               & 47616.4 & 11-14-1894 & 03-31-1989 & 13.472
               & 0.463 \\ 
g & ZTF        & 54 & 58107.4
               & 58482.5 & 12-20-2017 & 12-30-2018 & 14.136
               & 0.288 \\   
g & ASAS-SN  & 937 & 58220.2
             & 58933.2 & 04-12-2018 & 03-25-2020 & 13.954 
             & 0.130  \\
g & Lulin/SLT & 313 & 58397.9
               & 58832.7 & 10-06-2018 & 12-15-2019 & 14.128
               & 0.089 \\
g & Pan-STARRS & 12 & 55593.4
               & 56737.3 & 02-01-2011 & 03-21-2014 & 14.261
               & 0.382 \\
V & ASAS-3    & 217 & 51874.2 
               & 54862.2 & 11-26-2000 & 01-31-2009 & 13.793 & 0.310  \\
V & ASAS-SN & 960 & 55957.4
               & 58450.6 & 01-31-2012 & 11-28-2018 & 13.860
               & 0.324  \\ 
V & AAVSO & 10890 & 58404.3
               & 58915.2 & 10-13-2018 & 03-07-2020 & 14.094
               & 0.134  \\ 
r & ZTF & 494 & 58073.3
               & 58476.4 & 11-16-2017 & 12-24-2018 & 14.283
               & 0.311 \\    
r & Lulin/SLT & 13786 & 58397.9 
            	& 58938.6 & 10-06-2018 & 03-30-2020 & 14.190
               & 0.099 \\
r & Pan-STARRS & 10 & 55940.4
               & 56709.3 & 01-14-2012 & 02-21-2014 & 14.109
               & 0.175 \\
i & Lulin/SLT & 309 & 58397.9
               & 58832.7 & 10-06-2018 & 12-15-2019 & 14.293
               & 0.067 \\   
i & Pan-STARRS & 13 & 55195.4
               & 56652.5 & 12-30-2009 & 12-26-2013 & 14.332
               & 0.292 \\
z & Pan-STARRS & 14 & 55283.3
               & 56641.5 & 03-28-2010 & 12-15-2013 & 
               13.311 & 0.336 \\   
y & Pan-STARRS & 15 & 55276.2
               & 56641.5 & 03-21-2010 & 12-15-2013 & 
               14.327 & 0.423 \\
W1 & WISE & 208 & 55298.7 & 58780.5 & 04-12-2010 & 10-24-2019 &
               13.338 &  0.249  \\   
W2 & WISE & 208 & 55298.7 & 58780.5  & 04-12-2010 & 10-24-2019 &
               13.280 &  0.273  
\enddata
\end{deluxetable}

\subsection{Optical and Infrared Light Curve Data}

ZTF is a northern-sky synoptic survey project ($\delta > -30^\circ$) carried out by the 1.2-m Samuel Oschin Telescope at the Palomar Observatory.  With a large field-of-view mosaic CCD camera (47~deg$^2$ with 1.0$\arcsec$ pixel scale), the Galactic Plane can be scanned once a night with g and/or r filter. For HO\,Pup, 54 and 494 good quality measurements were taken in the g- and r-bands, respectively, between 2017 December and 2018 December. 

ASAS-SN data are taken by a couple of quadruple telescopes located in both hemispheres. The mounted cameras have a 4.5-deg$^2$ field of view with a pixel scale of 7.8$\arcsec$.  This survey provides us the longest-time baseline from 2012 to 2020 containing 960 V-band and 937 g-band measurements.  

The AAVSO responded quickly to get involved with the monitoring of HO\,Pup with V-band, right after our report of this extraordinary event to the ZTF community. On 2018 October 13, the AAVSO began to collect data with extremely good coverage in the time domain, using a ML16803 CCD camera with a pixel scale of 2.06$\arcsec$ equipped on a 0.4-m telescope at the Remote Observatory Atacama Desert \citep[ROAD;][]{ham12} in San Pedro de Atacama, Chile. Every clear night, ROAD monitored HO\,Pup more than 5 hours with an average of 50 measurements, resulting a total of 10890 photometric data points. The typical uncertainty of the measurements is about 0.02 mag. Here we used 0.15 mag as our quality threshold after testing several values.  

Pan-STARRS used five filters\footnote{For simplicity, we refer to them as g, r, i, z and y in the rest of this paper.}, g$_{p1}$, r$_{p1}$, i$_{p1}$, z$_{p1}$ and y$_{p1}$, to survey the sky. It has a wide-field 7-deg$^2$ mosaic camera with a pixel scale of 0.26$\arcsec$, mounted to the dedicated 1.8-m Pan-STARRS telescope, which is located at the Haleakale Observatory in Hawaii. HO\,Pup was observed over a full five-year time span (2009-2014). In contrast to ASAS-SN or AAVSO, there are only a small number of observations. Two significant dips ($m>$15.5 mag) were also recorded by Pan-STARRS.

The 0.4-m SLT telescope, located at the Lulin Observatory in Taiwan, was used to perform near-simultaneous gri-band and intensive r-band monitoring in follow-up observations of HO\,Pup. Together with the equipped Apogee U42 CCD, SLT images has a pixel scale of 0.79$\arcsec$. We used 77 reference stars located within 0.5$^\circ$ from HO\,Pup to perform differential photometry and calibrated to the Pan-STARRS catalog. The SLT data provides significant support to the variability investigation in the short-time scale and monitoring the color variation in the optical as the multi-band data from the optical surveys mentioned earlier were not taken simultaneously or nearly simultaneous (hence no color information).

WISE is an all-sky survey mission that mapped the entire sky at 3.4, 4.6, 12, and 22 $\mu$m (hereafter referred as W1, W2, W3, and W4) with spatial resolutions of 6.1$\arcsec$ and 6.4$\arcsec$ in W1 and W2 bands \citep{mai11}, respectively. The near-earth object WISE (NEOWISE) project is particularly designed to hunt for asteroids using the W1 and W2 bands, and it provides infrared (IR) data over nine years from 2010 April to 2019 October. The cadence of the WISE observations is about twice a year; each observation includes about a dozen of measurements over one day. 

Finally, we considered the utility of the DASCH and ASAS-3 light-curve data, but they were not used in subsequent analysis. The DASCH project has digitized photometric measurements from nearly 500,000 glass plates across 100 years. With a quality cut of 0.2 mag, 260 Johnson B-band magnitudes were selected for HO\,Pup from 1894 up to 1989 as historical records. One possible dip is included (see subsection \ref{sec:dips}). In case of ASAS-3 data \citep{poj97,poj02}, we excluded them in the analysis because of the associated large pixel scale ($\sim15\arcsec$). Therefore, issue of blending is unavoidable at the location of HO\,Pup. Nevertheless, we extracted 217 V-band light curves (grade A and B only) from the ASAS-3 archive. These light-curve data were measured using the smallest ASAS-3 aperture size \citep[][2 pixels; corresponding to the {\tt MAG\_0} in ASAS-3 catalogs]{poj05}. 

\subsection{Spectroscopic Data}

Soon after the 2.5-mag dips found by ZTF in late 2017, we collected spectra of HO\,Pup based on the observations carried out in 2018 by P60/SEDM and BOAO/BOES, in 2019 by CFHT/ESPaDOnS, as well as in 2020 by P200/DBSP. Due to the nature of queue observations for these telescopes and instruments, none of the spectra were taken during the 2.5-mag dip event of HO\,Pop, which did not occur again after mid-2018. 

The Spectral Energy Distribution Machine \citep[SEDM,][]{ben12,rit14,bla18,rig19} is a low-resolution IFU (integral field unit) spectrograph, mounted on the robotic P60 telescope at the Palomar Observatory \citep{cen06}, providing efficient follow-up observations.  The dispersion on the red side and blue side are 35 and 17.4~\AA~ per pixel, respectively. Queued observations of SEDM were carried out multiple times between 2018 October 19 and 2018 November 07, but only spectra from four nights were usable. The data were automatically reduced using the dedicated SEDM reduction pipeline \citep{rig19}. The low-resolution P60/SEDM spectra can only be used to identify H$\alpha$ emission lines, hence they were excluded in this work. 

\begin{figure*}
 \gridline{
    \fig{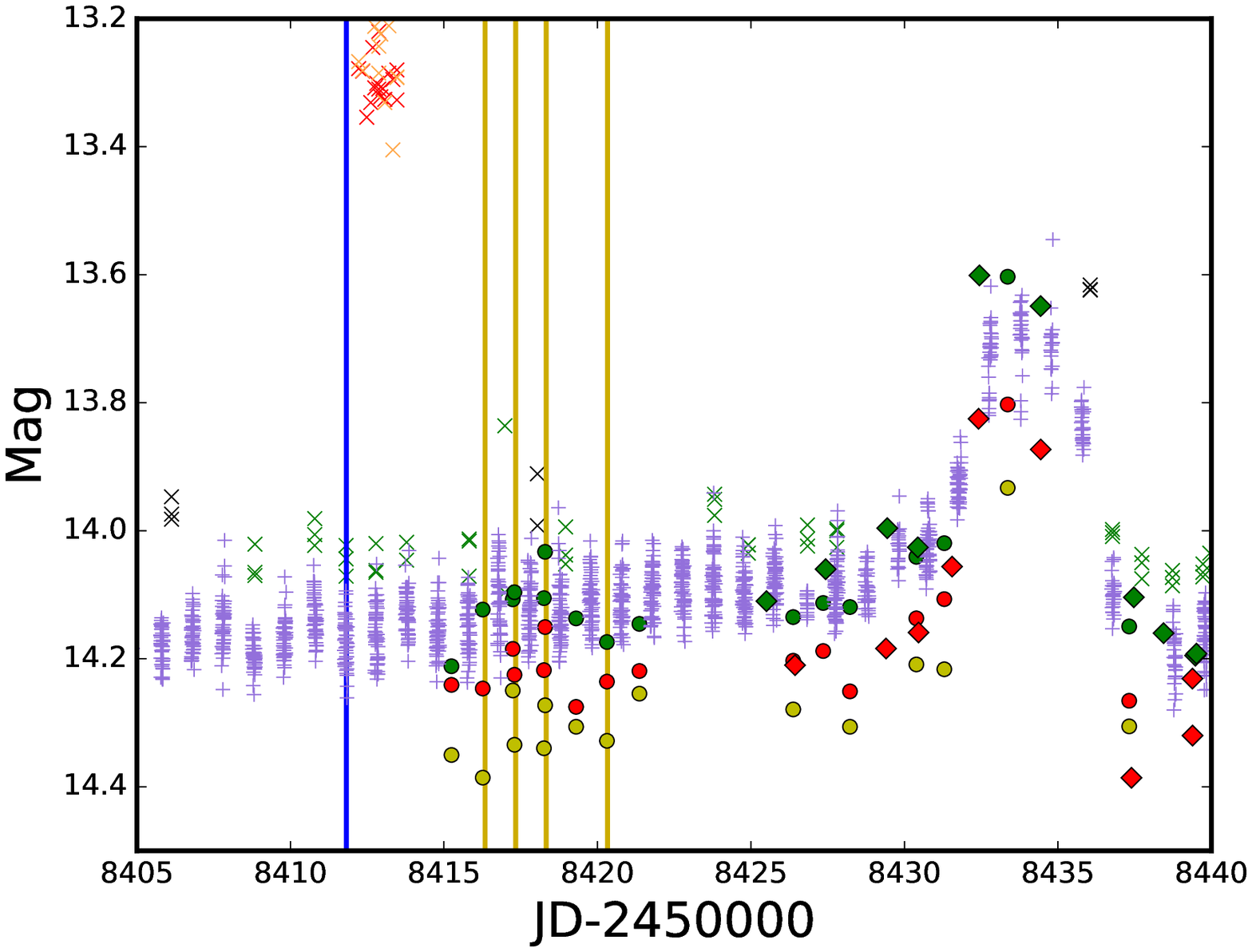}{0.475\textwidth}{}
    \fig{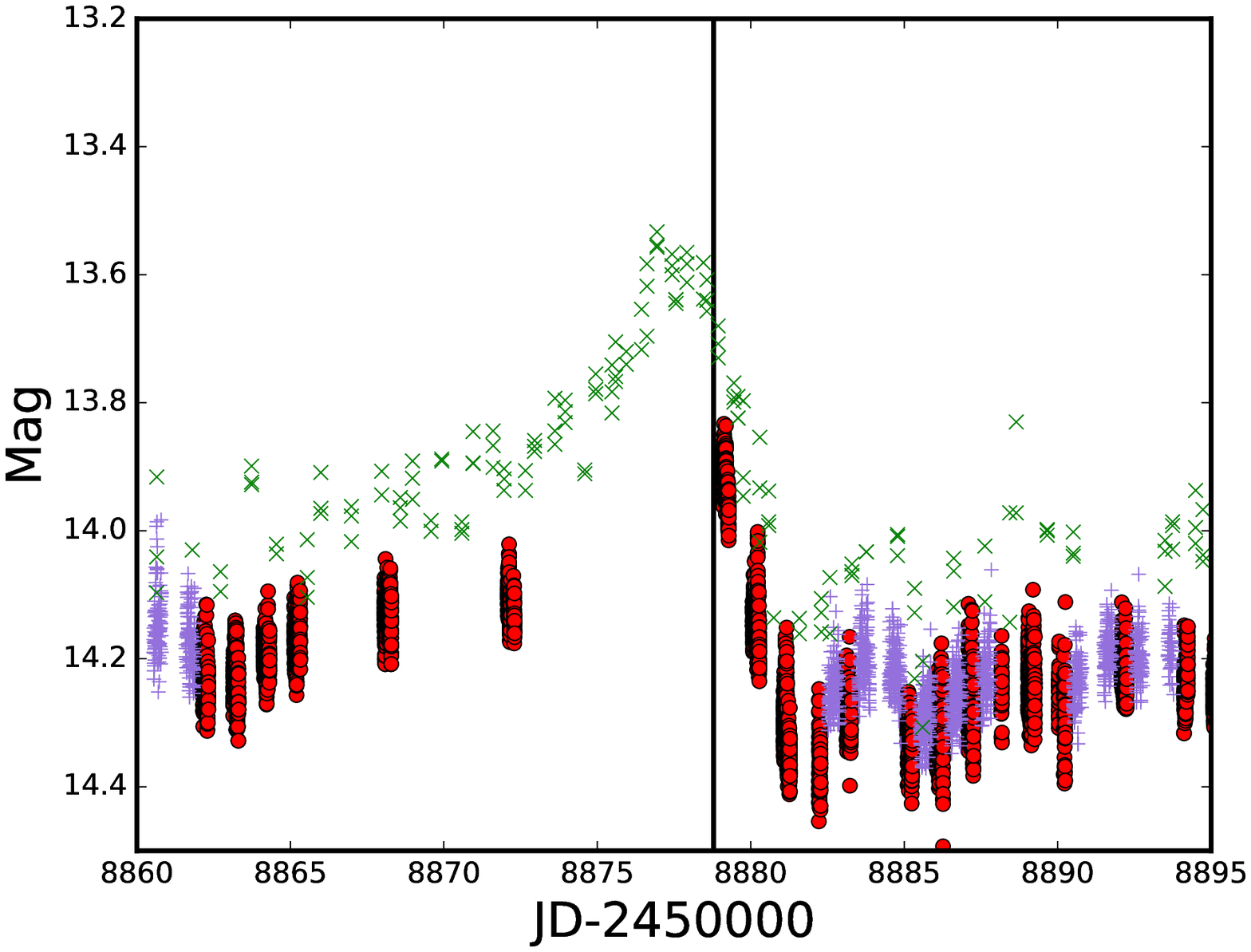}{0.475\textwidth}{}
    }
 \gridline{
    \fig{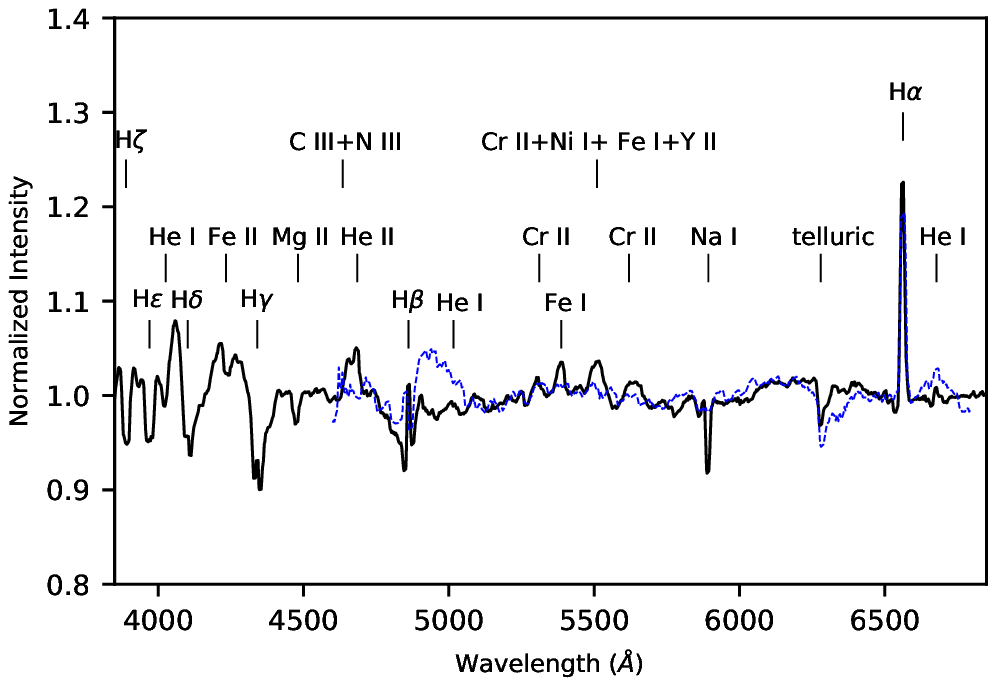}{0.975\textwidth}{}
    }    
\caption{The top panels are same as Figure~\ref{allLC}, except for the epoch with BOAO/BOES (blue line) and LOT/TRIPOL2 (yellow lines, see Table~\ref{pol}) observations on the top-left panel. The top-right panel is for the epoch with P200/DBSP spectroscopic observation (black line), taken right after the photometric maximum of one of the brightening event.  The bottom panel shows the BOAO/BOES spectrum (dashed blue curve) and P200/DBSP spectrum (solid black curve) resulted from a normalization using a $5^{\mathrm{th}}$ order polynomial. Both of the H$\alpha$ and H$\beta$ lines are obviously seen in these spectra with either low or medium resolution, hence the emission nature of HO\,Pup was essentially confirmed.  The spectral characteristics at the phase of brightening, including hydrogen Balmer lines, as well as helium, iron, nickel, sodium and chromium lines, are identified. 
}
 \label{5m}
\end{figure*}

We have also obtained an optical spectrum of HO\,Pup using the Bohyunsan Optical Echelle Spectrograph \citep[BOES,][]{kim02} in long-slit mode, mounted on the 1.8-m Optical Telescope at the Bohyunsan Optical Astronomy Observatory (BOAO) in South Korea, on 2018 October 20. The observation was conducted using the grating 300V with 3$\arcsec$ slit width under 1.8$\arcsec$ seeing, giving a spectral resolution of $R\sim$1200. The wavelength coverage is 3500--7000~\AA~and the exposure time is 40 minutes. We follow the standard procedures of data reduction using IRAF\footnote{IRAF is distributed by the National Optical Astronomy Observatory, which is operated by the Association of Universities for Research in Astronomy, Inc., under cooperative agreement with the National Science Foundation.}, e.g., flat-fielding, wavelength calibration with FeNeArHe lamp, and flux calibration with the standard star G191-B2B. The normalized spectrum is shown in the bottom panel of Figure~\ref{5m}. The H$\alpha$ emission line is clearly detected with an  equivalent width (EW) of $\sim - 4.3$~\AA, estimated using the Astrolib PySynphot package \citep{sts13}. Due to low signal-to-noise ratio below 5000~\AA~of the spectrum, we cannot confirm the detailed spectral type of HO\,Pup. 

\begin{figure*}
 \centering
 \includegraphics[width=0.495\columnwidth, angle=0]{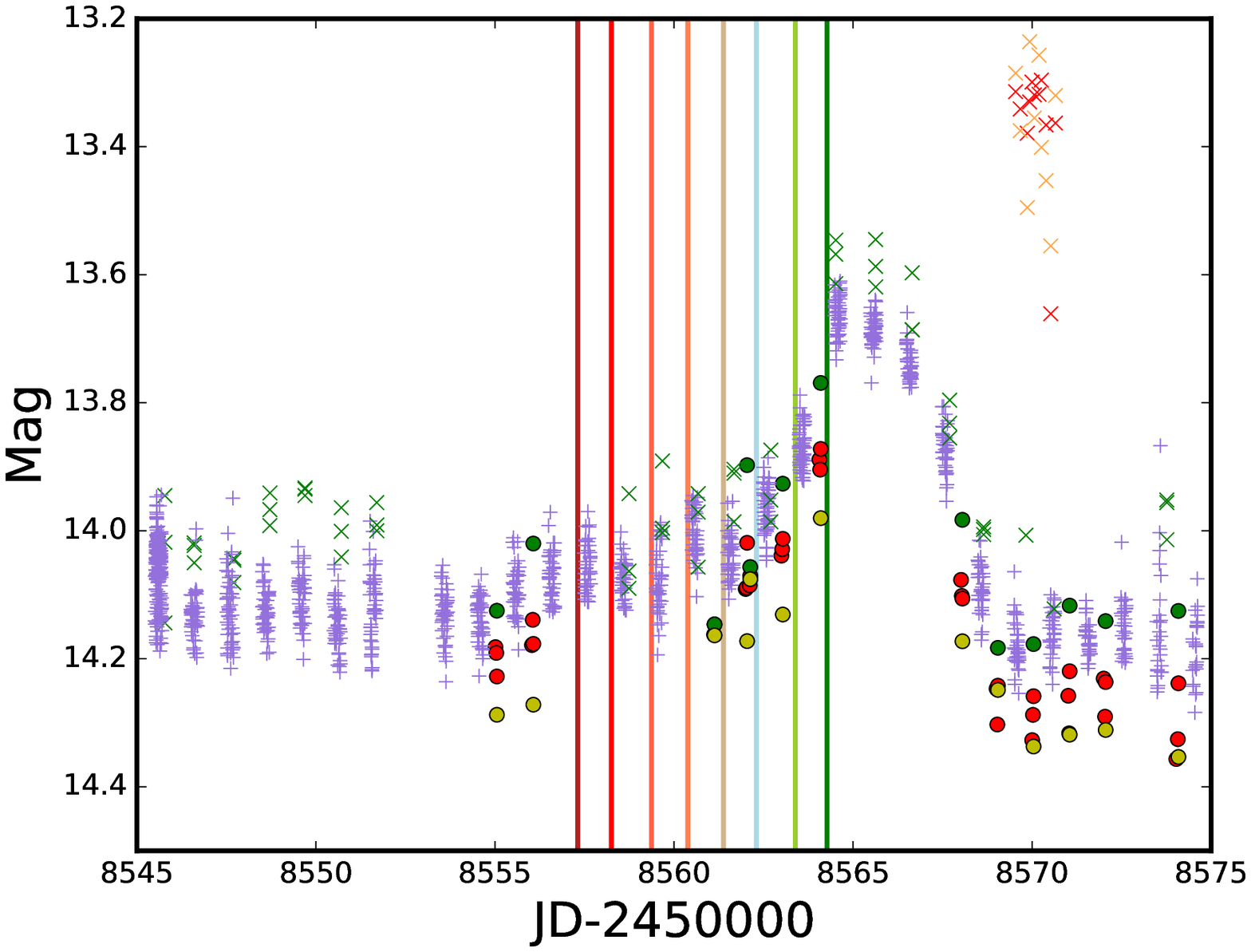}\\
 \includegraphics[width=0.5\columnwidth, angle=0]{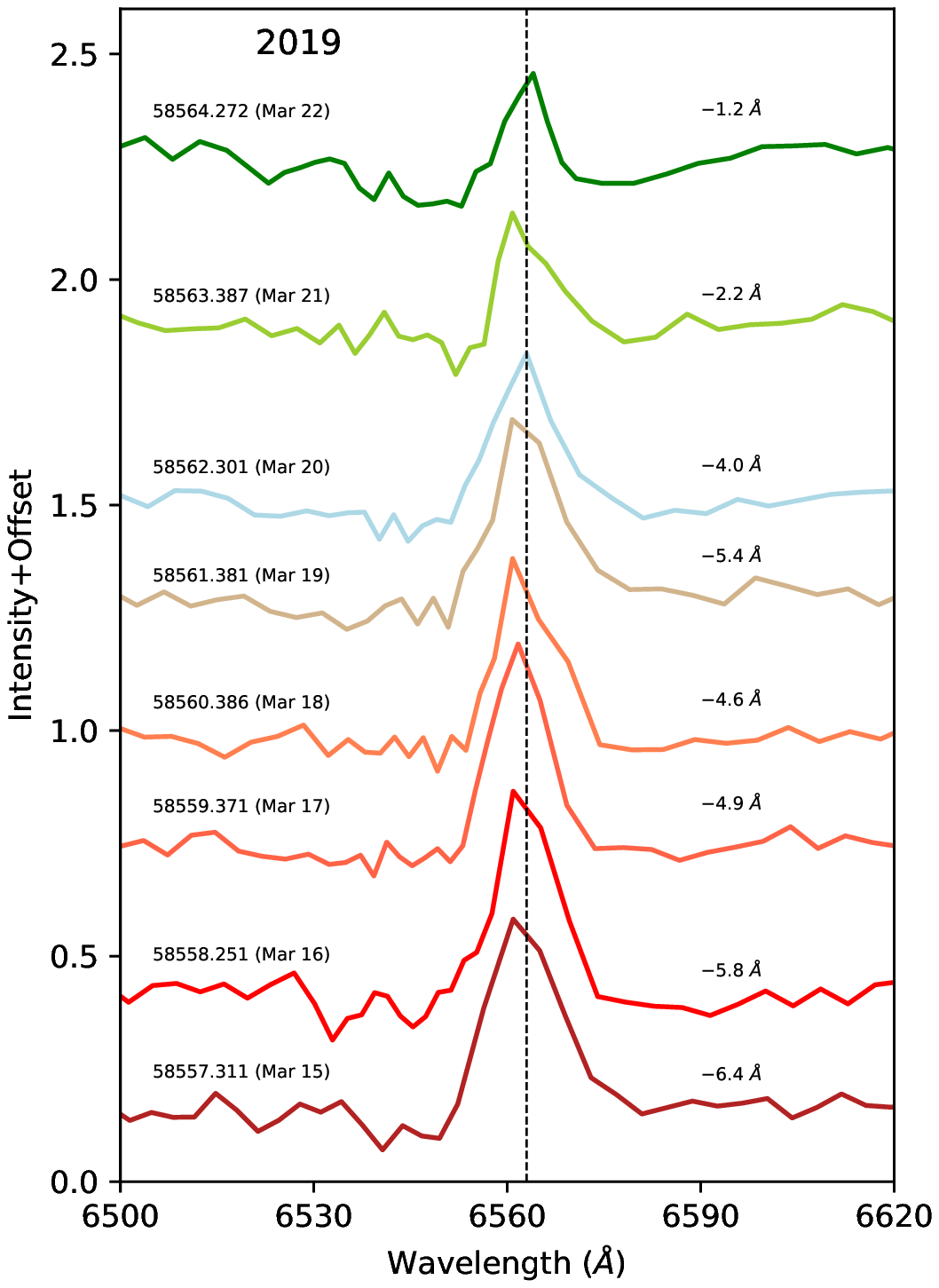}
\caption{The top panel is same as Figure~\ref{allLC}, except for the epoch with CFHT/ESPaDOnS observations. The bottom panel shows the CFHT/ESPaDOnS spectra of HO\,Pup taken from 2019 March 15 to 22, centered on the H$\alpha$ line (vertical dashed line). To improve the signal-to-noise ratio in the plot, the spectra were re-binned with a resolution of $\delta\lambda \sim 3$\AA. Values on the right of each spectra are the measured EW of the H$\alpha$ line, derived using the Astrolib PySynphot package \citep{sts13}.}
 \label{CFHT}
\end{figure*}

To further investigate the H$\alpha$ emission variability, we took high-resolution echelle spectra of HO\,Pup using the ESPaDOnS (Echelle SpectroPolarimetric Device for the Observation of Stars) mounted on the 3.6-m Canada-France-Hawaii Telescope (CFHT), with a spectral resolution of $R\sim$ 68,000. The wavelength coverage is 3750$-$10400 \AA~at approximately 0.007$\mbox{\AA}$/pixel, containing 40 grating orders.  With an exposure time of 2400s for spectroscopy (sky+star mode) and 4800s for spectropolarimetry (both were arranged by the CFHT queue observers), HO\,Pup was nightly observed from 2019 March 15 to 2019 March 22 under good seeing ($\lesssim$1\arcsec) conditions. CFHT provided a set of fully reduced spectra from Libre Esprit, an automatic ESPaDOnS reduction package/pipeline \citep[][]{don97,don07}. One of the brightenings or brightening events from 14.1 mag to 13.6 mag was fortunately well observed during our observing runs. As shown in Figure \ref{CFHT}, all H$\alpha$ emission lines are clearly seen, despite that the parts of the continuum were barely observed.    

Finally, a single 600-second spectrum was obtained using the Palomar 200-inch Hale Telescope (P200) with the Double-Beam Spectrograph \citep[DBSP,][]{oke82} on 2020 January 30. A $2\arcsec$ slit was used with the 600/4000 grating in the blue arm and the 316/7500 grating in the red arm, providing $R\sim1500$. A wavelength calibration spectrum was taken directly after the science spectrum. A spectrum of a flux-standard star was obtained at the beginning and end of the night. The data were reduced using the pipeline developed by \citet{bellm2016}\footnote{https://github.com/ebellm/pyraf-dbsp}. Using the pipeline, we performed the standard bias corrections, flat-field corrections, wavelength calibration and flux-calibration. We also automatically combined the red and blue spectra using this pipeline. The reduced spectrum is displayed in the lower panel of Figure \ref{5m}, in which the H$\alpha$ EW value is $\sim - 4.0$~\AA, also estimated using the Astrolib PySynphot package.

\subsection{Polarimetric Data}

\begin{deluxetable}{l CCC C}
\tabletypesize{\footnotesize}
\tablecaption{Polarization values of HO\,Pup  \label{pol}
			}
\tablecolumns{7}
\tablewidth{0pt}
\tablehead{ 
	\colhead{Date} & 
		 \colhead{$P_r$} & \colhead{$\theta_r$} & \colhead{$P_i$} & \colhead{$\theta_i$}\\
	\colhead{} & \colhead{(\%)} &  \colhead{($\degr$)} & \colhead{(\%)} &  \colhead{($\degr$)} 
          }
\startdata
2018 October 24 & 
               0.7 \pm 0.5 &  44.4  \pm 22.5  & 
               0.9 \pm 0.2 &  45.8  \pm 34.1  \\	
2018 October 25 &
               0.5 \pm 0.5 &  135.8 \pm 28.6  &
               0.5 \pm 0.5 &  132.9 \pm 28.6  \\ 
2018 October 26 & 
               1.5 \pm 0.2 &  28.1  \pm 11.3  & 
               0.8 \pm 0.3 &  30.8  \pm 42.4  \\
2018 October 28 & 
               0.0 \pm 0.5 &  25.6  \pm 28.6  & 
               0.7 \pm 0.7 &  128.3 \pm 28.6  \\     
\enddata
\end{deluxetable}

In addition to the photometric and spectroscopic observations, we also took polarimetric data by using the TRIPOL2 (second generation of the Triple Range Imager and POLarimeter) instrument \citep{sat19}, equipped on the Lulin One-meter Telescope (LOT), in 2018 October. TRIPOL2 can simultaneously take polarization images in Sloan g'-, r'- and i'-band with the half-wave plate rotating to four angles: $0\arcdeg$, $22\arcdeg.5$, $45\arcdeg$ and $67\arcdeg.5$.  We measured the flux at each angle using aperture photometry following standard reduction procedures, and the Stokes parameters (I, Q and U) were then derived.  The polarization percentage $P=\sqrt{Q^2+U^2}/I$ and the position angle $\theta=0.5\arctan{(U/Q)}$ can be calculated from these Stokes parameters with a typical accuracy of $\Delta P\lesssim0.3\%$.  Polarimetric data of HO\,Pup was well obtained in r- and i- bands (but it was too faint in the g-band), as reported in Table~\ref{pol}.  We have also observed a number of unpolarized and polarized standard stars \citep{sch92} to calibrate the instrumental polarization and angle offset \citep{sat19}. Based on the observations of these standard stars, we found that the performance of TRIPOL2 was very stable within those nights for both of the measurements of polarization percentages (using both polarized and unpolarized standard stars) and the polarization position angles (using the unpolarized standard stars); for more details, see \citet{hua19}.  To reduce the influence by sky conditions, we also require that each of these four measurements has at least five sets of images with nearly the same count in each angles.  This ensures us the polarization measurements are reliable from our nightly observations. 

\section{RESULTS} \label{sec:r&a}

\subsection{An Emission-Line Object} 

As mentioned in the Introduction, HO\,Pup could be considered as a Be star candidate or an IW And-type DN, without any emission-line or even spectral observations reported in the literature. With our spectroscopic follow-up observations, the H$\alpha$ emission line is clearly seen in each of our observations including the spectra from BOAO/BOES, CFHT/ESPaDOnS and P200/DBSP (as shown in Figure~\ref{CFHT} and \ref{5m}).  The H$\alpha$ emission EW is weak, e.g. $\sim -4.5$\AA~as measured in BOAO/BOES spectrum. Therefore, in addition to the known GCAS or IW\,And type light-curve variation, HO\,Pup is now confirmed as an emission-line object.

Beyond H$\alpha$ lines, other prominent hydrogen, helium, and metal features can be easily identified. During the brightening phase, HO\,Pup shows Balmer lines, from H$\beta$ to H$\zeta$, which are superposed on weak emission lines (see Figure~\ref{5m}). The helium features show absorption lines \ion{He}{1} 4026~\AA~and 4471~\AA, and emission lines \ion{He}{1} 5016~\AA, 6678~\AA~and \ion{He}{2}~4713~\AA.  Several metal absorption features can be identified, such as \ion{Fe}{2} 4233~\AA, \ion{Mg}{2}4481~\AA, \ion{Na}{1} doublet 5889~\AA~and 5893~\AA; other emission features include \ion{Cr}{2} 5311~\AA~and 5620~\AA, and \ion{Fe}{2} 5387~\AA. We also identified a few blended emission lines such as \ion{C}{3}+\ion{N}{3} 4634-4651~\AA~(blended Bowen lines) and \ion{Cr}{2}$+$\ion{Ni}{1}$+$\ion{Fe}{1}$+$\ion{Y}{2} 5510~\AA.  Note that the spectral characteristics and EW variation of H$\alpha$ emission for HO\,Pup will be discussed further in Section~\ref{sec:dis}.

\subsection{The Dip Events}\label{sec:dips}

\begin{figure}
 \centering
 \includegraphics[width=1\columnwidth]{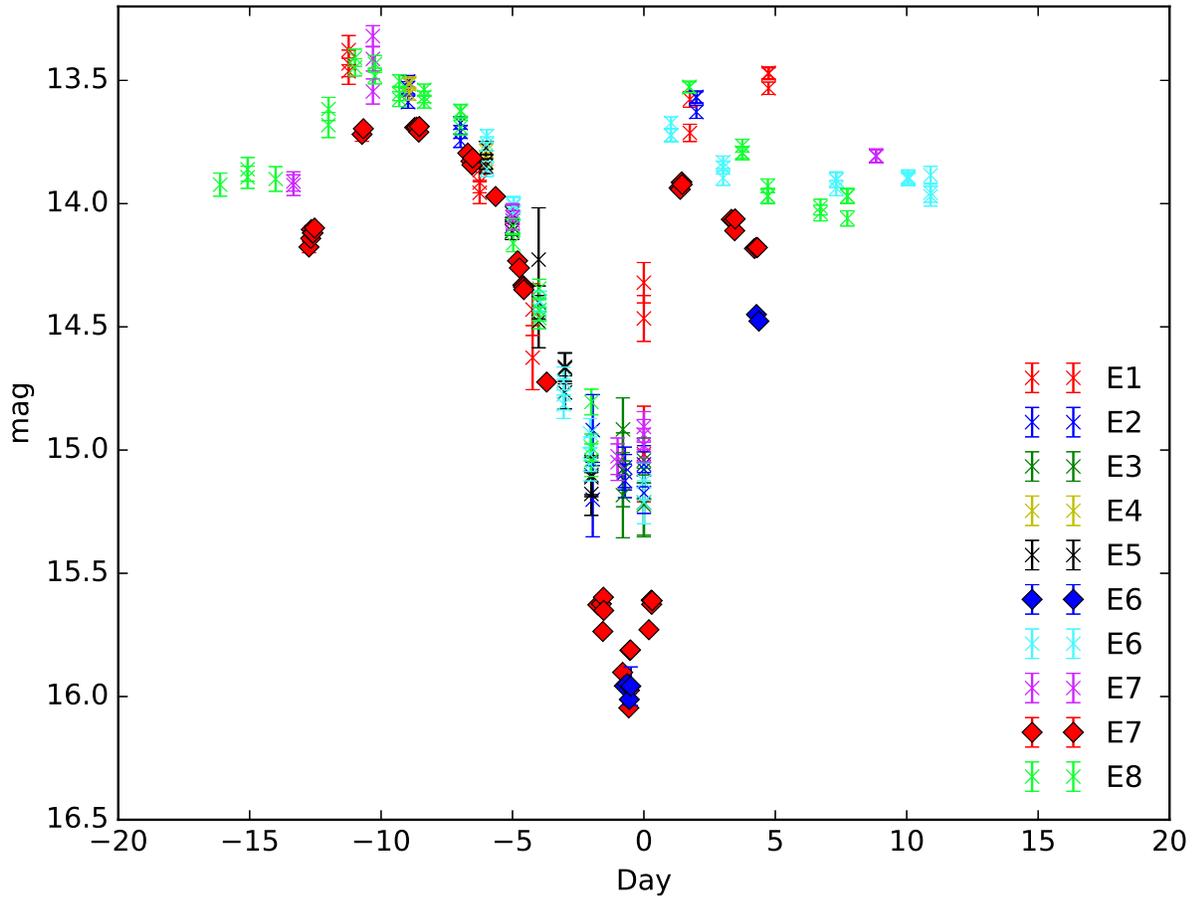}
	\caption{The dip events observed in the 2017-2018 season, where the diamonds and crosses represent the ZTF and ASAS-SN data, respectively.  Two 2.5-mag dips, E6 and E7, are over-plotted with six other similar dips, including E1, E2, E3, E4, E5 and E8.  We aligned these events using their own fading slope, which have almost the same decreasing trends.  With visual inspection, the minimum of the dips are chosen as the day of zero in the $x$-axis. 
	}
 \label{2017dips}
\end{figure}

Two 2.5-mag dips were observed in the ZTF r and g bands (13.5-16 mag; see Figure~\ref{hopupztf}) in late 2017. During these significant events, ASAS-SN also witnessed these two dips in the V band.  These dips observed in both independent surveys are well matched in the time sequence. When examining the long-term light curve taken from ASAS-SN, in total there are eight dip events found between 2017 September and 2018 January, labeled as E1 to E8 in Table~\ref{dips}. Note that the dip event E4 was barely observed by ASAS-SN on 2017 October 23 (JD$=$2458049.806). Within half a day, WISE happened to catch the IR counterpart of the same event down to 14.58 mag in W2 band (JD$=$2458049.435). Just like other dips in optical bands, the IR brightness drops significantly at this event when compared to the average brightness in WISE W2 band (13.280 mag).  

The E1-E8 dips have very similar shapes, especially their decreasing slopes. In Figure~\ref{2017dips}, we over-plotted these events all together to show their similar shapes. Note that only the E6 and E7 events were found to drop by $\sim 2.5$-mag, while the other events show the drop in magnitudes range from $\sim 1$ to $\sim 2$~mag. This could be due to the issue of sampling the light curve around these events (i.e., the observations miss the epochs of the exact event minimum). However, these dip events aperiodically occurred in the range of $\sim7$ to $\sim30$ days.  Using the entire light-curve coverage from E1 to E8, the most likely event duration was found to be around 14.3 days estimated using phase dispersion minimization \citep[PDM,][]{ste78} method.

In the early epochs from 2011 to 2015, the dips B2 to B5 were individually recorded by Pan-STARRS or ASAS-SN, as seen in Table~\ref{dips}. A suspected dip, B1, was also recorded in the DASCH light curve around 1934. In addition, one more dip, A1, was caught by ZTF during the 2018-2019 season. Before A1 occurred (2018 November 20), we have four successful SEDM observations. Unfortunately, there is a long queue of observing requests on SEDM until the end of our proposed observation run on 2018 November 27. Thus, we did not have any further observation and missed the opportunity to catch spectroscopic observations of this dip.  

\begin{deluxetable}{l CcC CCCC}
\tabletypesize{\footnotesize}
\tablecaption{Faintest measurements of each dip event.   \label{dips}
			}
\tablecolumns{9}
\tablewidth{0pt}
\tablehead{
	\colhead{JD} & \colhead{Date} & \colhead{Surveys} & \colhead{Mag} & \colhead{error}  & \colhead{Band} & \colhead{Event No.} 
		 \\
          }
\startdata
2427525.315 & 1934/03/28 & DASCH     & 15.170 & 0.150 & B & B1\\
2455645.785 & 2011/03/25 & Pan-STARR & 15.119 & 0.005 & y & B2 \\
2456016.802 & 2012/03/30 & ASAS-SN & 15.526 & 0.07 & V & B3 \\
2456596.153 & 2013/10/30 & Pan-STARR & 15.646 & 0.006 & y & B4 \\
2457038.010 & 2015/01/15 & ASAS-SN & 15.576 & 0.156 & V & B5 \\
2458007.137 & 2017/09/10 & ASAS-SN & 15.017 & 0.194 & V & E1 \\
2458021.849 & 2017/09/25 & ASAS-SN & 15.175 & 0.084 & V & E2 \\
2458033.880 & 2017/10/07 & ASAS-SN & 15.224 & 0.122 & V & E3 \\
2458049.806 & 2017/10/23 & ASAS-SN & >15.285 &  & V & E4 \\
2458049.435 & 2017/10/23 & WISE & 14.581 & 0.174 & W2 & E4 \\
2458056.777 & 2017/10/30 & ASAS-SN & 15.178 & 0.087 & V & E5 \\
2458074.053 & 2017/11/16 & ZTF & 15.958 & 0.078 & r & E6 \\
2458074.771 & 2017/11/17 & ASAS-SN & 15.211 & 0.088 & V & E6 \\
2458101.029 & 2017/12/13 & ZTF & 15.975 & 0.01 & r & E7 \\
2458101.081 & 2017/12/13 & ASAS-SN & 15.05 & 0.074 & V & E7 \\
2458130.982 & 2018/01/12 & ASAS-SN & 15.041 & 0.066 & V & E8 \\
2458442.964 & 2018/11/20 & ZTF & 15.5 & 0.087 & g & A1 \\   
\enddata
\end{deluxetable}

\subsection{Semi-Regular 0.5-Magnitude Brightening Events} \label{sec:0.5m}

\begin{figure}
    \centering
    \includegraphics[width=1\columnwidth]{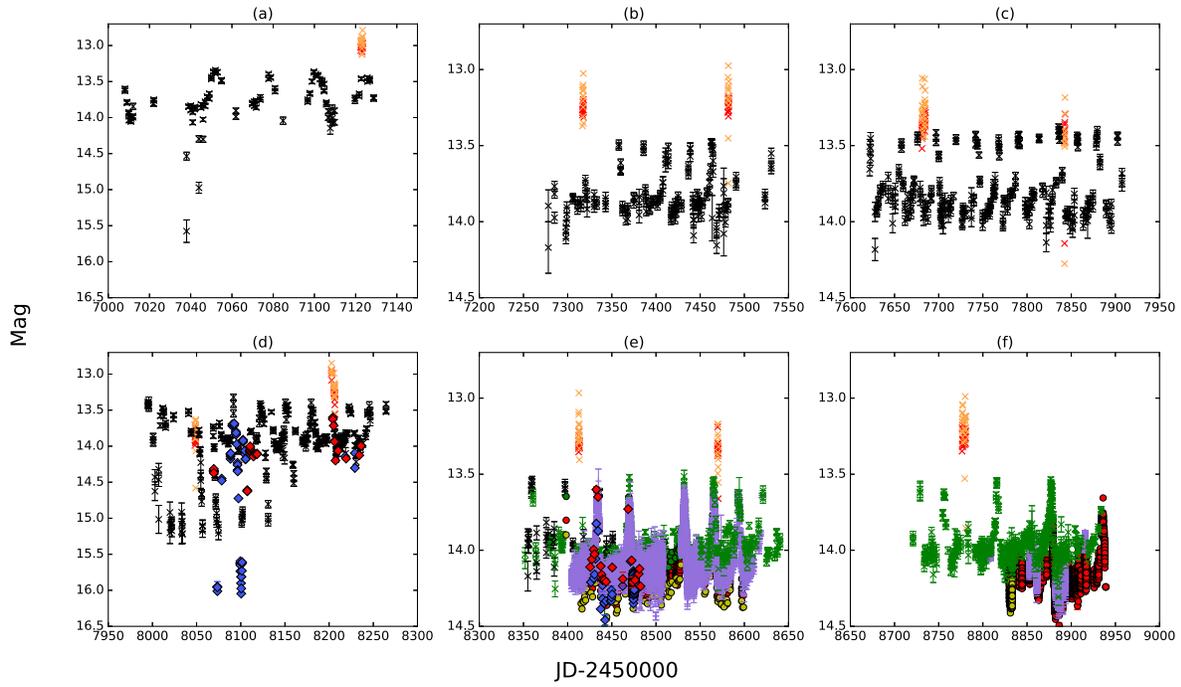}
    \caption{Same as Figure~\ref{allLC}, but time scale enlarged for the different years of interest: (a) 2014-2015, (b) 2015-2016, (c) 2016-2017, (d) 2017-2018, (e) 2018-2019 and (f) 2019-2020.
    }
    \label{smallarea}
\end{figure}
    
In addition to the dip events mentioned in the previous subsection, the long-term light curves of HO\,Pup also exhibit semi-regular $\sim 0.5$~mag brightening events known as brightenings. These brightening events can been seen in Figure~\ref{allLC}, for example, during the 2016 and 2017 seasons, which are better visualized in upper panels of Figure~\ref{5m}. In Figure~\ref{smallarea}, we split the $\sim7$~years of light curves of HO\,Pup into different segments. The $\sim 0.5$~mag brightening events are clearly showing in the light curves repeatedly with a duration between 21 to 61 days, as displayed in Figure~\ref{smallarea} (a) to (f).

\begin{figure}
    \centering
    \includegraphics[width=1\columnwidth]{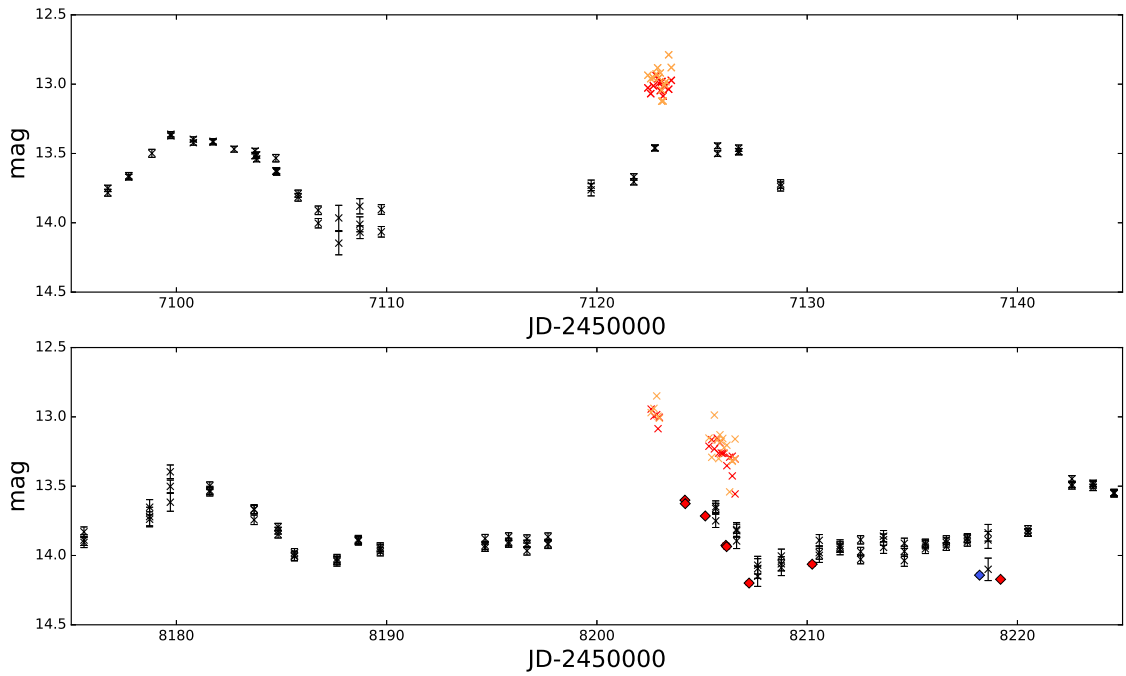}
    \caption{Same as Figure~\ref{allLC}, but time scale enlarged for the two $\sim0.5$~mag brightening events as observed with the WISE data in 2015 (upper panel) and 2018 (lower panel).}
    \label{WISE_brightening}
\end{figure}

An initial analysis suggested that these brightening events exhibit some periodicity but no single period can be determined. Therefore, we estimated the periodicity on different segments of the light curves using the PDM approach. These segments are similar to those presented in Figure~\ref{smallarea}. We derived a mean period of 25.9~days and 22.8~days for the light-curve data in 2015-2016 (JD: 2457250 - 2457500) and 2016-2017 (JD: 2457500 - 2457950), respectively.  For 2017-2018 (JD: 2457950 - 2458300), since the light curves are dominated by the dip events as shown in the previous subsection, we excluded this part of the light curves for our PDM analysis. In 2018-2019 (JD: 2458300 - 2458650), visual inspection suggested that the periodicity of the brightening events varied sporadically with longest and shortest periods of 37 and 26 days, respectively. Therefore we divided the light curves of 2018-2019 into three parts (in JD): 2458300 - 2458485, 2458485 - 2458570 and those latter than 2458570. The corresponding periods found by applying the multi-band PDM analysis are $35.6\pm0.03$, $31.9\pm0.18$ and $26.8\pm0.34$~days, respectively. Finally, in the recent epoch 2019-2020 (JD: 2458720 - 2458933), the estimated period turns out to be $60.9\pm0.28$~days.  It is worth pointing out that these repeatable brightening events resemble the light curve of IW\,And-type stars \citep{kim20}.  

In addition to the brightening events in the optical, two brightening events were also observed in the mid-IR from WISE. As shown in Figure \ref{WISE_brightening}, these brightening events in the mid-IR were almost varying together with the optical ASAS-SN V-band light curve, suggesting the entire continuum rises up during the brightening events.

For cases in which spectroscopic observations are possible, we have continuously taken spectra every night using CFHT/ESPaDOnS before the optical light curve reached its maximum brightness in one of the brightening events that occurred in 2019 March (see Figure~\ref{CFHT}). We found that the strength, or equivalently the EW, of H$\alpha$ line decreases while HO\,Pup became brighter, as shown in the spectra of the last two days in Figure~\ref{CFHT}.  

\subsection{Color Variations}

\begin{figure}
    \centering
    \includegraphics[width=1\columnwidth]{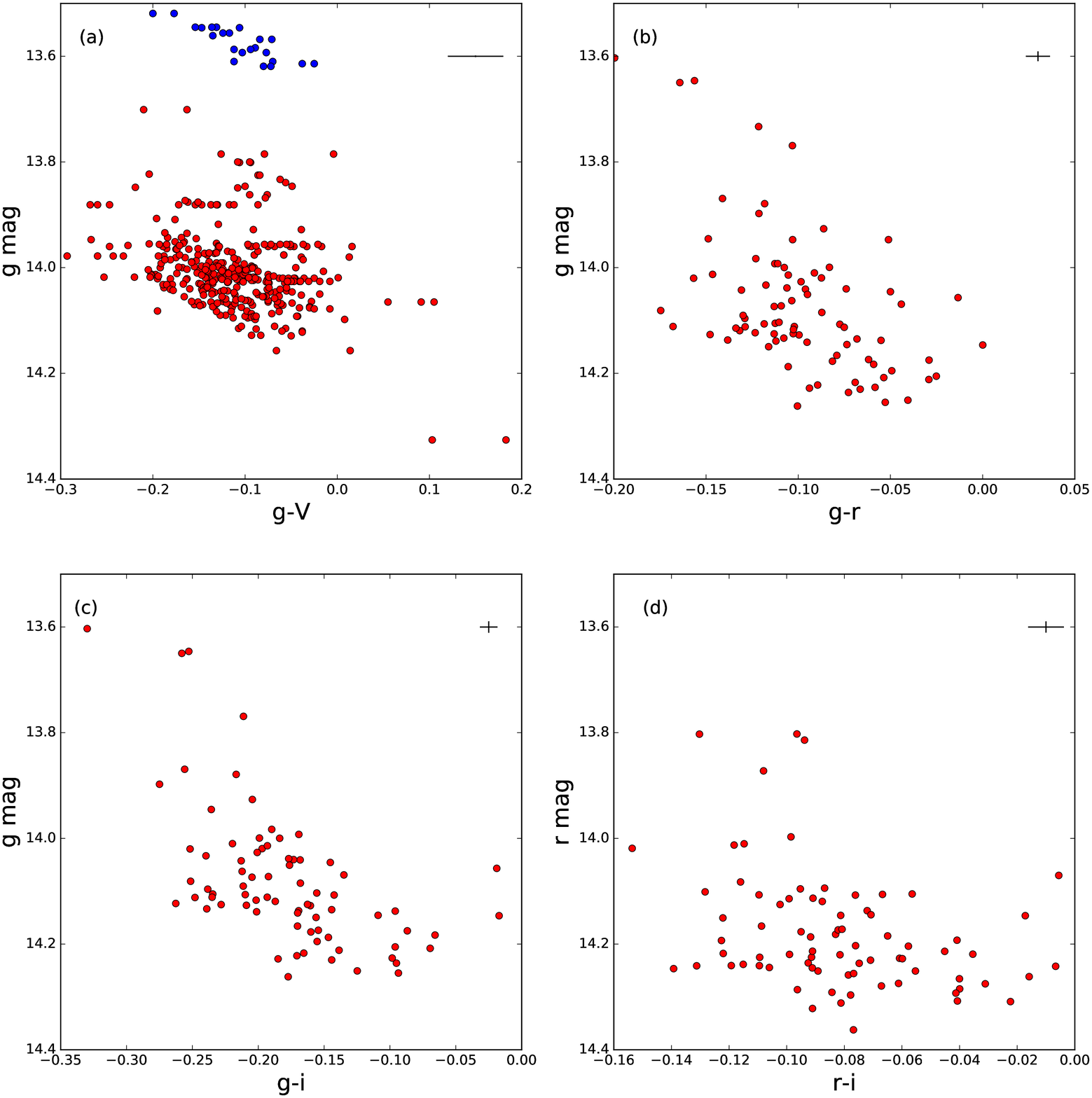}
    \caption{The color-magnitude diagrams (CMD) of HO\,Pup. The representative error bars of colors and magnitudes are shown in the upper right corners of these four CMDs.  The g- and V-band data in panel (a) were taken from ASAS-SN and AAVSO, respectively. The intense observations from both surveys allow us to separate the data points for those that occurred at the $\sim0.5$~mag brightening events (blue points) as discussed in sub-section \ref{sec:0.5m}, and those outside these events (red points). In panel (b) to (d), the gri-band data are based on the SLT observations taken at the Lulin Observatory. All of these CMDs show that HO\,Pup becomes bluer when it gets brighter. 
    }
    \label{color-mag}
\end{figure}

We investigated the color variations of HO\,Pup using several measurements. To construct the $(g-V)$ color, we selected g-band data from ASAS-SN and V-band data from AAVSO. Each g-band point and its corresponding V-band point were observed within 5 minutes. Also, we used gri-band data of the SLT telescope at the Lulin Observatory to construct the $(g-r)$, $(r-i)$ and $(g-i)$ colors, where the gri-band images were always taken nearly simultaneously, e.g.,within 5~minutes. To quantify the color variation with brightness, we plotted the magnitudes as a function of color (i.e. the color-magnitude diagram, CMD) in Figure~\ref{color-mag}. In all these optical colors, the blue data points were those corresponding the $\sim0.5$~mag brightening events and the red data points were taken at the quasi-standstill phase. As can be seen from panel (a) of Figure~\ref{color-mag}, the colors at these two phases cover a very similar range. Furthermore, all of the colors present a trend such that colors become bluer when HO\,Pup becomes brighter, especially for the $(g-i)$ colors shown in panel (c) of Figure~\ref{color-mag}.

\subsection{Variabilities at Short Time Scale}

\begin{figure}
    \centering
    \includegraphics[width=1\columnwidth]{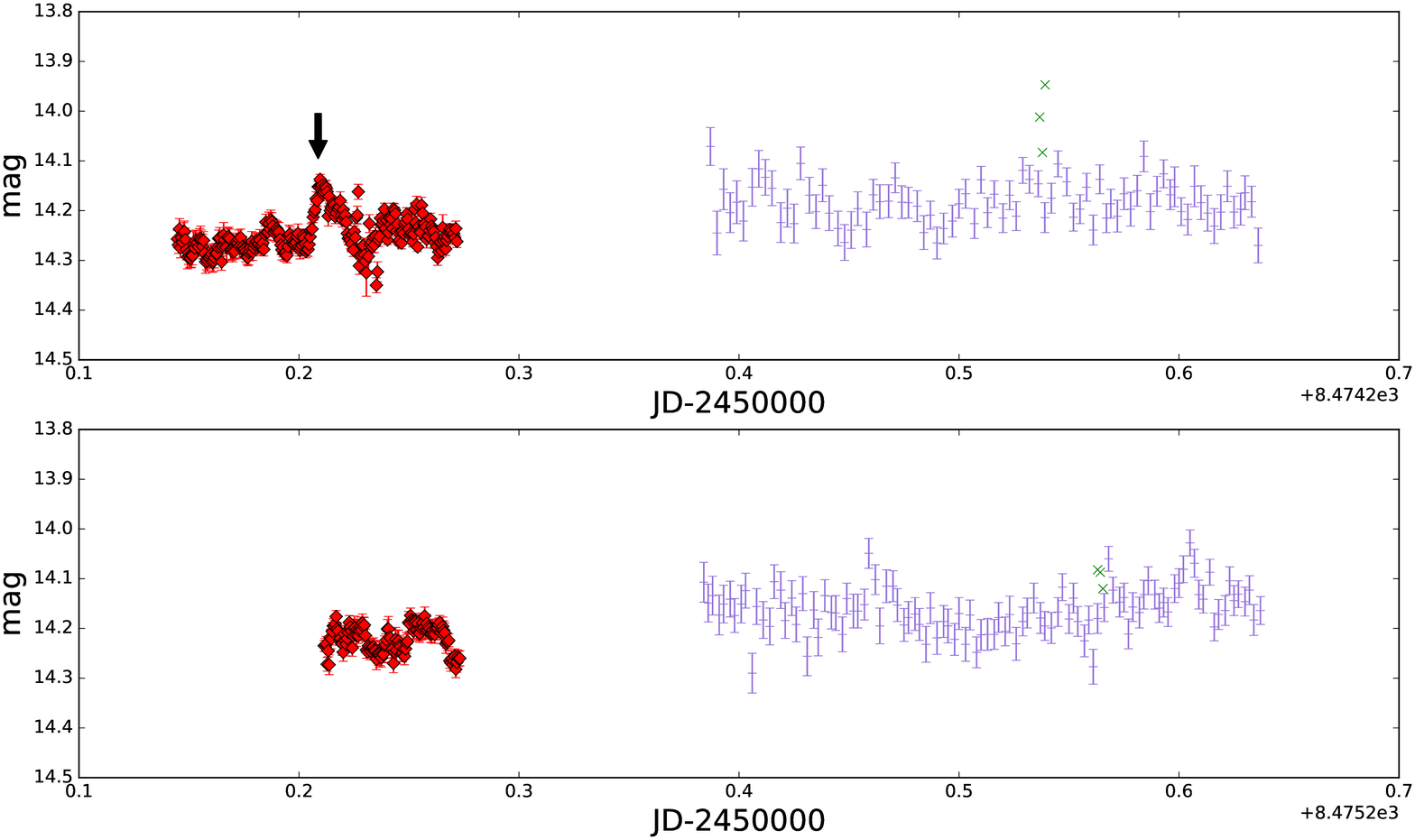}
    \caption{Same as Figure~\ref{allLC}, but with the time scale covering 0.6 days. Based on the ZTF data taken on 2018 December, some variabilities in the timescale of hours are clearly observed. On December 22 (upper panel), we see a flipped S-shaped event (indicated by an arrow) lasting about an hour. On the next day, December 23 (lower panel), the ZTF light curve obtained over two hours shows a variability with sinusoidal-like profile. Even with lower cadence or photometric precision, similar variations are also seen in the AAVSO light curve. 
    }
    \label{hour_scale}
\end{figure}

\begin{figure}
    \centering
    \includegraphics[width=0.9\columnwidth]{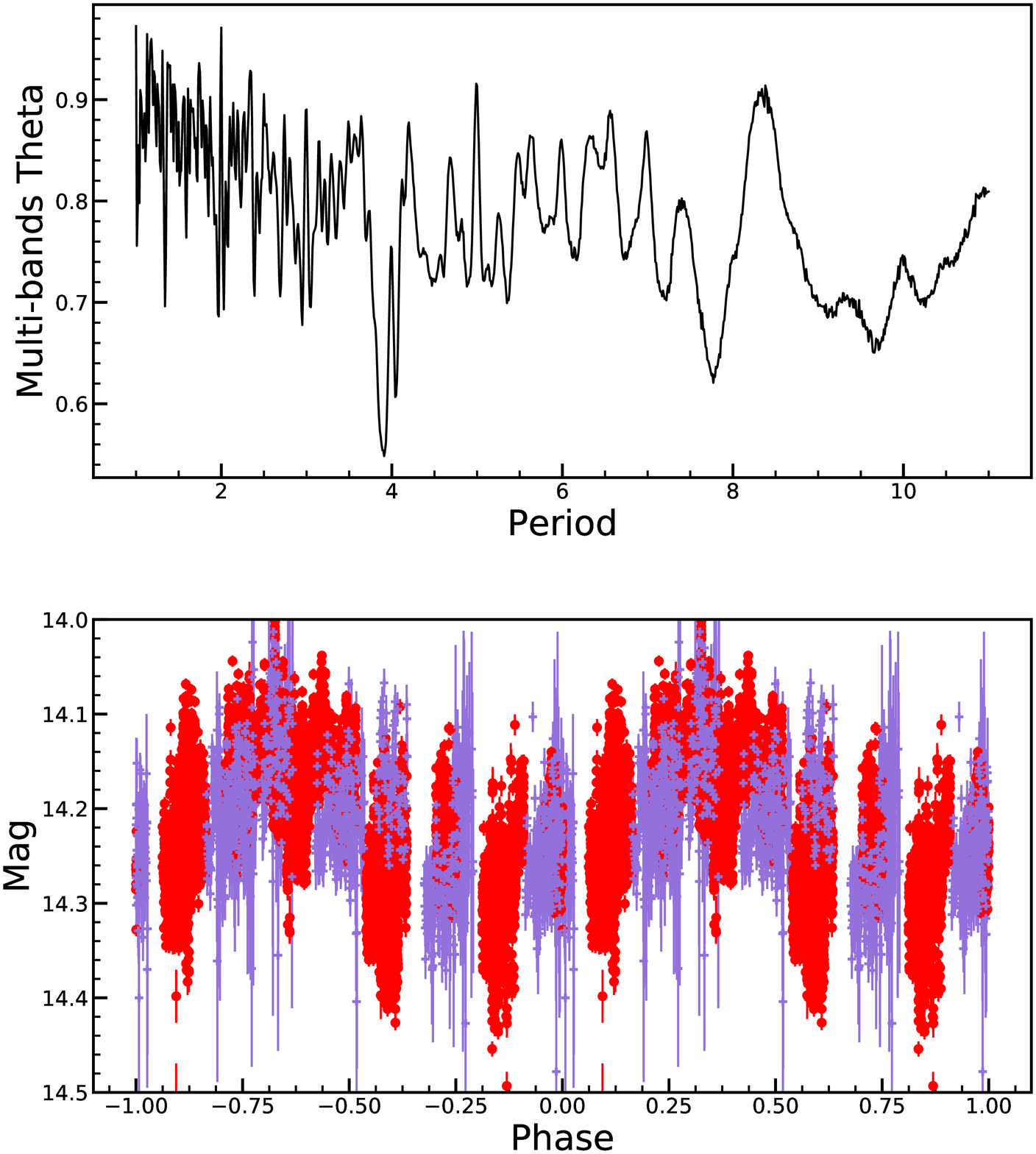}
    \caption{Multi-band PDM analysis of the oscillatory variation at the timescale of days. A period of $3.9\pm0.005$~days is clearly detected (upper panel) from light curves based on the superposition of two independent observations. The bottom panel is the phased light curves folded with the detected 3.9-day period, in which the symbols are same as in Figure~\ref{allLC}, revealing a sinusoidally shaped light curve. 
    }
    \label{days_scale}
\end{figure}

In addition to the dip events and brightening events, as presented in sub-section \ref{sec:dips} and \ref{sec:0.5m}, respectively, short-term variabilities were also found in the light curves of HO\,Pup during its quasi-standstill phase. For example in late December of 2018, ZTF performed continuous cadence observations on the Galactic Plane that included HO\,Pup. Therefore, continuous r-band light curves taken within 2 hours were available from ZTF on 2018 December 22 and 23, with 280 and 136 measurements, respectively (see Figure~\ref{hour_scale}). An hour-scale sinusoidal variability with $\sim0.1$~mag to $\sim0.2$~mag amplitude was revealed in this set of ZTF light curve. Similar variability can also be seen from the AAVSO light curve, even though the cadence of AAVSO data is not as high as ZTF in these two nights. On December 22, we also observed a transient event with a flipped S-shape profile. Due to the insufficient ($\sim2$~hours) time span of the ZTF light curves on these variability, we do not perform periodicity analysis for these two nights.  

\begin{figure}
    \centering
    \includegraphics[width=1\columnwidth]{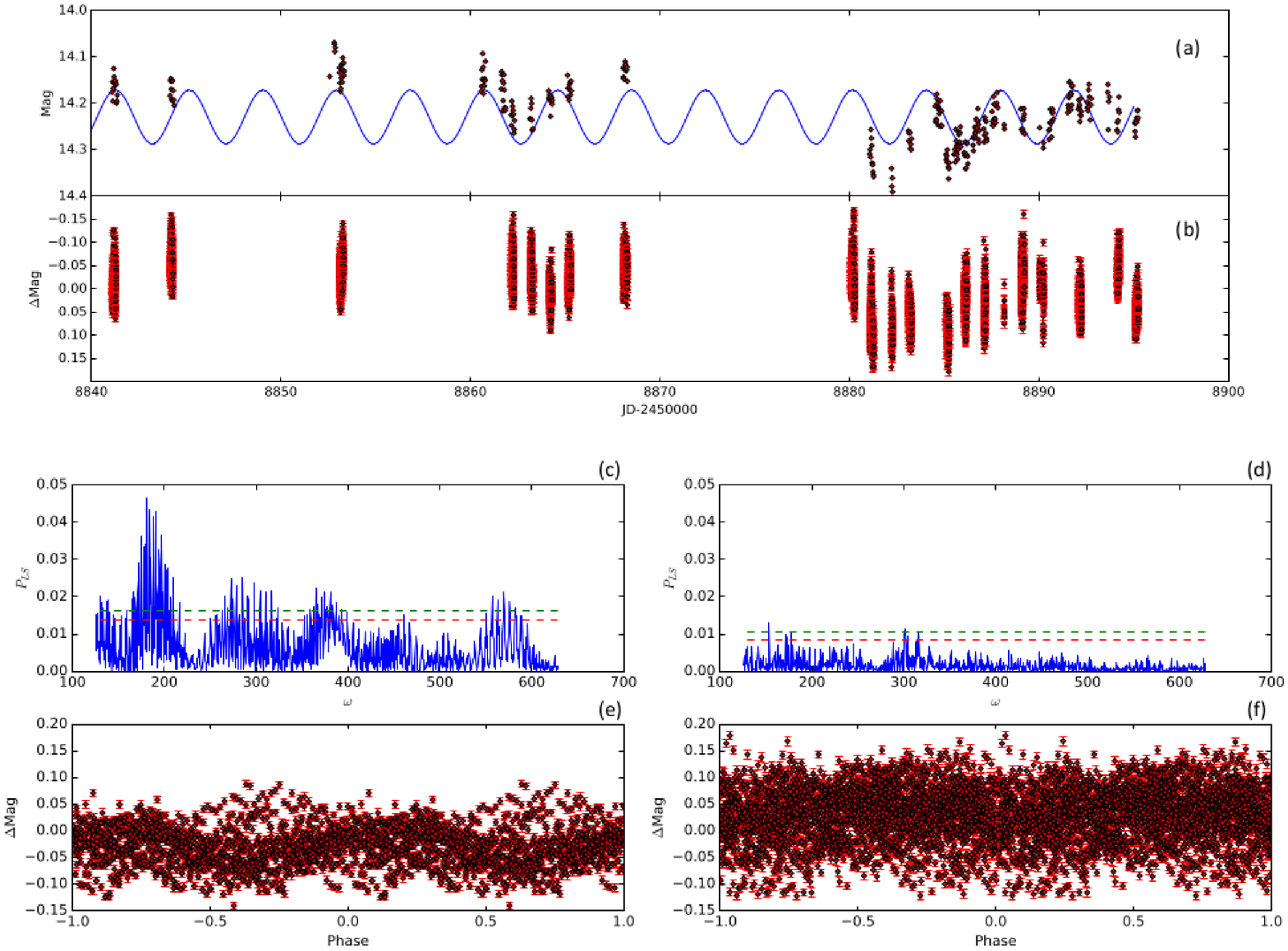}
    \caption{The variability of HO\,Pup down to timescale of hours. Panel (a) presents the SLT r-band photometric data taken from 2019 December to 2020 March. The blue curve represents the 3.9-day cycle fitted with a sinusoidal function. (see Figure~\ref{days_scale}).  To uncover the variability shorter than one day, this light curve has been subtracted with the sinusoidal function of 3.9-day and the residuals are displayed in panel (b). After removing the 3.9-day cyclic variations, we found a new cycle of 50-minute variations using Lomb-Scargle periodogram analysis, based on the light curve data taken from 2019 December to 2020 January, as shown in panel (c). As can be seen from this panel, the variation signal is well above the false-alarm probability (FAP, indicated by the red and cyan dashed lines for the 95$\%$ and 99$\%$ FAP, respectively). The variability with 50-minute period can be easily seen in the phased light curve as presented in panel (e).  However, the 50-minute variations were barely resolved or undetected for the data collected between 2020 February and March, as evident in panel (d) and (f). Instead, complicated variations can be seen in the nightly chunks of light-curve data within this period of observations.}
    \label{50minutes}
\end{figure}

To verify the short timescale variability of HO\,Pup light curve, we performed high-cadence monitoring of HO\,Pup with the SLT telescope in late 2019 to 2020. These high-cadence observations were done in a single r-band filter with 30-s exposure time for 2 to 5 hours continuously in available clear nights. Using both PDM and Lomb-Scargle periodogram routines, the resolved periodicity are 3.9 days and 50 minutes, respectively, as seen in Figure~\ref{days_scale} and \ref{50minutes}. Data taken around the brightening (for 5 nights) and about an hour before the dawn were removed in order to maintain high-quality light curves to detect the short timescale variability. In addition to the 3.9-day and 50-minute variabilities, two very weak periods, at 20 minutes and 90 minutes, respectively, were also detected but cannot be confirmed. 

\subsection{Polarization Variation}

We have also observed variations in polarization for HO\,Pup. Based on polarization level $P$ and position angle $\theta$ measured from Lulin Observatory (see Table~\ref{pol}), a small and yet significant polarization was observed on 2018 October 24 and 26.  However, the polarization drops back to an insignificant value on 2018 October 25 and 2018 October 28.  The observed variation in polarization suggested HO\,Pup exhibits intrinsic polarization.

\section{DISCUSSION} \label{sec:dis}

\subsection{Spectral-Energy-Distribution Fitting}\label{sec:class}

\begin{figure*}[ht]
 \gridline{
    \fig{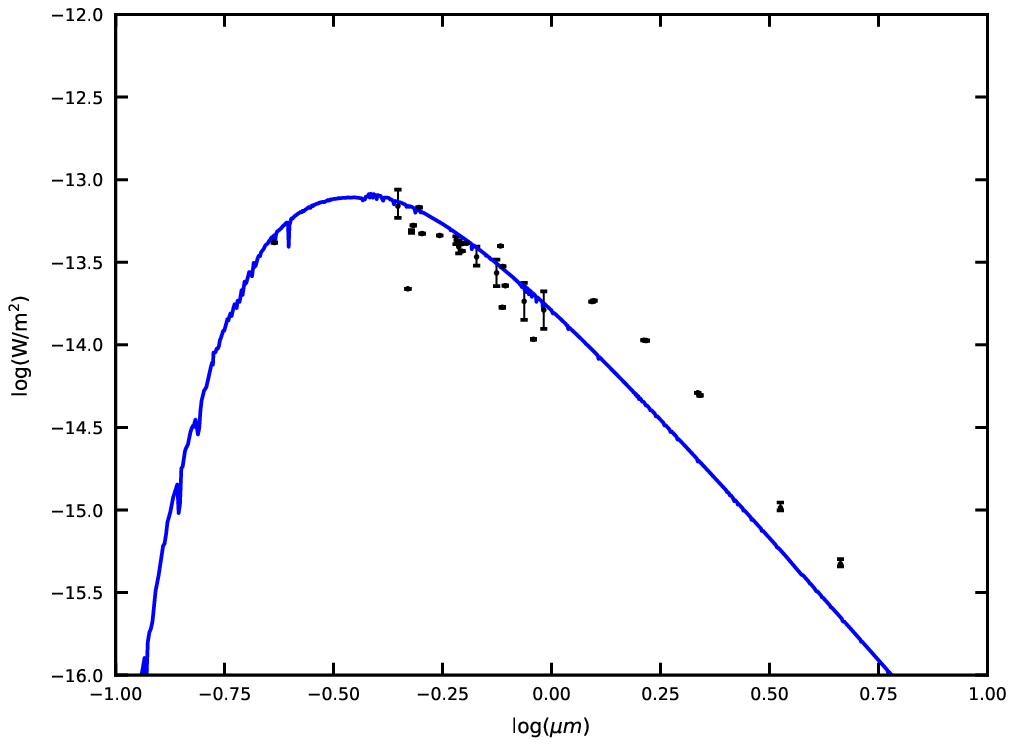}{0.53\textwidth}{(a) Fitting with a Be star spectrum at $16$~kpc}
    \fig{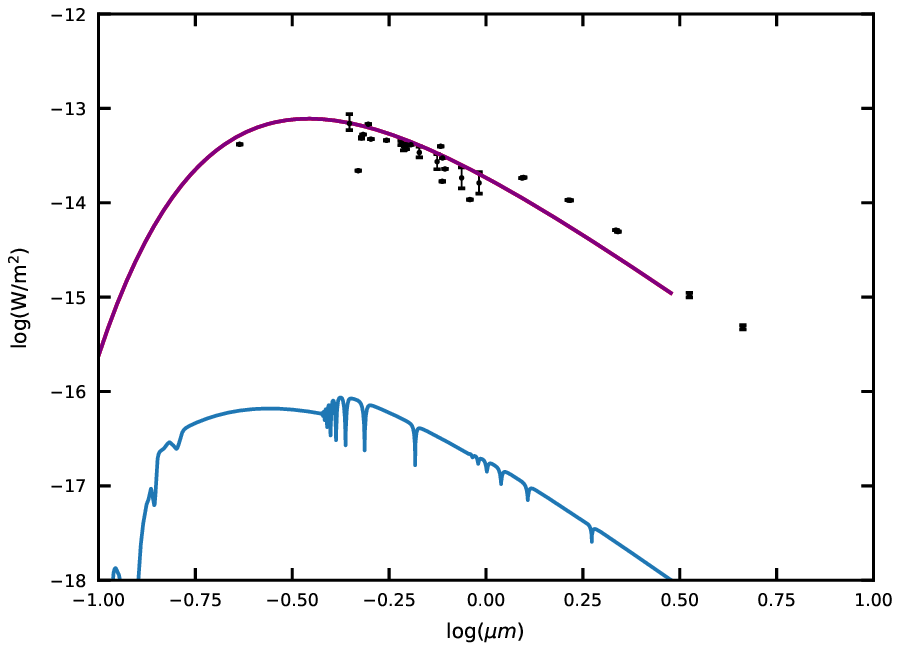}{0.53\textwidth}{(b) Fitting with a hot disk black-body at $618$~pc}
    }
\gridline{ 
    \fig{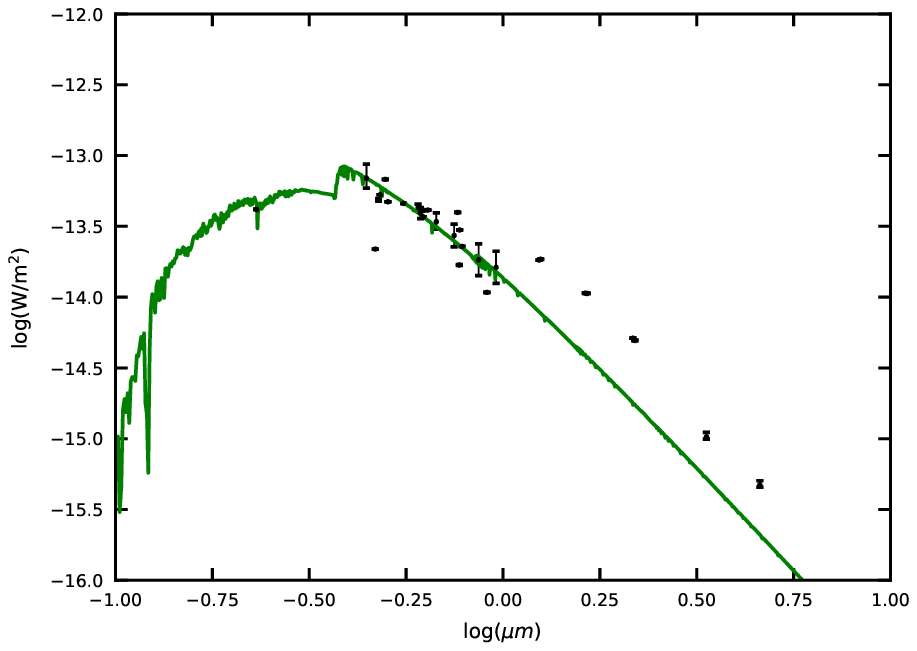}{0.53\textwidth}{(c) Fitting with a hot sub-luminous star spectrum at $618$~pc}
    }
    \caption{Fitting of the observed SED of HO\,Pup (black points from broadband photometry; see Table~\ref{sed_data}) with various theoretical spectra. In panel (a), the blue curve is the theoretical spectrum for a B1V dwarf, taken from the \citet{kur93} stellar atmosphere model, at a distance of $16$~kpc with $A_V=0.5$. In panel (b), the light blue curve is a representative theoretical white dwarf spectrum (with effective temperature of 12,000~K) adopted from \citet{tb09} and \citet{koe10}, available at the Spanish Virtual Observatory, at a distance of 618~pc with $A_V=0.1$. The purple curve represents the sum of the white dwarf spectrum and a hot disk with radius of $0.5$ Solar radius assuming black body radiation at 11,000~K, both at the same distance of 618~pc with $A_V=0.1$. In panel (c), the green curve is a theoretical spectrum for a hot sub-luminous star with an effective temperature of 12,000~K, taken from the \citet{kur93} stellar atmosphere model, assuming a distance of $618$~kpc with $A_V=0.1$.
    }
    \label{SED}
\end{figure*}

Prior to our work, classification of HO\,Pup in the literature was done via inspection of its light curves, as either a Be star with GCAS-type variability \citep{sam17} or as an IW And-type DN \citep{kim20}.  Here, we first estimated its spectral and luminous class using spectral-energy-distribution fitting together with distance estimation based on the Gaia-Data-Release-2 (DR2) parallax measurement \citep[$1.6165\pm0.0311$~mas, or $d=618.6$~pc,][]{gai18}. The observed SED of HO\,Pup was constructed using available broadband photometry covering from near-UV (NUV) to mid-infrared. In Table~\ref{sed_data} we listed all available broadband photometric data of HO\,Pup queried within 1$\arcsec$ radius using the {\tt VizieR} photometry viewer \citep{och00}.  As shown in the Figure~\ref{SED}, the SED peaked between the NUV and B-bands, which indicates it is either a hot early-type star or a source with a hot component.

Assuming HO\,Pup is a luminous B-type main-sequence star, fitting these broadband photometric data with a theoretical SED model \citep{kur93} suggested that the spectral class of HO\,Pup has to be B1V, as demonstrated in Figure~\ref{SED}(a), with the corresponding extinction of $A_V = 0.5$ and at a distance of 16~kpc. However, the distance of 16~kpc is inconsistent with the Gaia DR2 distance. If the Gaia DR2 distance is correct, we then derived a $M_{V}\sim4.7$~mag for HO\,Pup by adopting an apparent magnitude of $m_{V}=13.7$~mag and an extinction of $A_V=0.1$~mag. This value of absolute magnitude $M_V$ is consistent with the known IW\,And-type stars, as summarized in Table \ref{Mv}. Therefore, the possibility of HO\,Pup being a Be star is ruled out.

As presented in previous sections, the IW And-type phenomenon can be seen in the light curves of HO\,Pup collected in our work \citep[as well as in][]{kim20}. Assuming HO\,Pup is a IW And-type DN, which, in general, consist of a white dwarf as the primary star, a (low-mass) secondary companion star, and a (hot) disk. In Figure \ref{SED}(b), a representative theoretical spectrum for a white dwarf was shown at a distance provided by Gaia DR2, $618$~pc, which has negligible contribution to the observed SED of HO\,Pup (the black points in Figure \ref{SED}). Assuming HO\,Pup has a hot disk with a temperature of 11,000~K and a radius of $0.5$ Solar radius, the black-body spectrum of such a hot disk can fit well to the observed SED of HO\,Pup at a distance of $618$~pc, as presented in Figure \ref{SED}(b). We believe this hot disk represents the ensemble average of the disk during the quasi-standstill phase.

The majority of the broadband photometric data listed in Table ~\ref{sed_data} did not have observation times available. Therefore, photometry at phases of both brightenings and quasi-standstills possibly could be included as part of the observed SED data.\footnote{For example, the B-band data ranged from $13.24-15.16$ mag (see Table~\ref{sed_data}) suggesting one of the brightest data points might be taken during the brightening phase.} Given that the observed flux of the SED peaked between the NUV-band and the B-band, it turns out that the observing time in the NUV band is the key to determining which phase that we fit in the observed SED, and it was found to be taken on 2012 February 02 (JD = 2455959.9603; according to the header information available from the corresponding GALEX image). Based on the two data points taken from ASAS-SN with $V\sim 13.9$~mag within 10~days\footnote{The next brightening event occurred on 2012 February 17, which were $\sim14$-day after the NUV data was taken. The $\sim14$~days duration between NUV observation and the next brightening was shorter than the typical duration of two consecutive HO\,Pup brightening events, which is longer than 20~days.} before and after the NUV observation at the quasi-standstill phase, it is very likely that the NUV observation was taken while HO\,Pup was at the quasi-standstill phase.

Alternatively, the observed SED of HO\,Pup can also be fitted with a hot sub-luminous star as its secondary, as shown in Figure \ref{SED}(c). We note that an early spectroscopic observation has suggested the prototype object, IW And, could be an early type dwarf or subdwarf \citep{mei80}.

\begin{deluxetable}{llllr}
\tabletypesize{\footnotesize}
\tablecaption{Collected broadband photometric data for HO\,Pup used in SED fitting.
\label{sed_data}	}		
\tablecolumns{4}
\tablewidth{0pt}
\tablehead{ 
	\colhead{Band} & \colhead{$\lambda (\mu$m)} &  
		\colhead{mag} & \colhead{error} & \colhead{reference} 
          }
\startdata
NUV       & 0.231  & 15.135 & 0.011 & \citet{mor07} \\
Bj        & 0.435  & 15.16  & 0.42  & \citet{las08} \\
B         & 0.444  & 14.233 & 0.57  & \citet{zac12} \\
B         & 0.444  & 13.8   & 0.43  & \citet{las08} \\ 
B         & 0.444  & 13.240 & $\cdots$      & \citet{zac05} \\ 
Bf        & 0.468  & 13.86 & 0.44  & \citet{las08} \\
g       & 0.481 & 14.079 & 0.556  & \citet{gei19}\\
g       & 0.481 & 14.237 & 0.192 & \citet{hei18}\\
g       & 0.481 & 13.818 & 0.273 & \citet{wol18}\\
g       & 0.481 & 14.199 & 0.204 & \citet{cha16}\\ 
$G_{BP}$  & 0.532 & 14.121 & 0.029 & \citet{gai18}\\ 
V       & 0.551 & 14.082 & 0.49 & \citet{zac12}\\
V       & 0.551 & 13.340 & $\cdots$ & \citet{zac05}\\
r       & 0.617 & 14.085 & 0.013 & \citet{hei18}\\
r       & 0.617 & 13.974 & 0.272 & \citet{wol18}\\ 
r       & 0.617 & 14.043 & 0.008 & \citet{cha16} \\
r       & 0.617 & 14.074 & 0.42 & \citet{zac12}\\
R       & 0.658 & 14.170 & $\cdots$ & \citet{zac05}\\
G       & 0.673 & 14.083 & 0.009 & \citet{gai18}\\
i       & 0.752 & 14.391 & 0.003 & \citet{wol18}\\
i       & 0.752 & 14.150 & $\cdots$ & \citet{hei18}\\
i       & 0.752 & 14.118 & $\cdots$ & \citet{cha16}\\
i       & 0.752 & 13.887 & 0.10 & \citet{zac12}\\
In       & 0.784  & 14.04  & 0.43  & \citet{las08}\\ 
$G_{RP}$  & 0.797 & 13.833 & 0.049 & \citet{gai18}\\
z       & 0.866 & 14.586 & 0.003 & \citet{wol18}\\
z       & 0.866 & 14.363 & 0.015 & \citet{hei18}\\
z       & 0.866 & 14.311 & 0.017 & \citet{cha16} \\
y       & 0.962 & 14.354 & 0.054 & \citet{hei18}\\
y       & 0.962 & 14.327 & 0.050 & \citet{cha16} \\
J       & 1.25 & 13.300 & 0.026 & \citet{sku06}\\ 
H       & 1.65 & 13.137 & 0.027 & \citet{sku06}\\
$K_s$     & 2.15 & 13.152 & 0.037 & \citet{sku06}\\
W1      & 3.40 & 13.480 & 0.028 & \citet{cut12}\\ 
W2      & 4.60 & 13.358 & 0.038 & \citet{cut12} \\
W3      & 12.00  & 12.564 & 0.471 & \citet{cut12} \\
\enddata
\end{deluxetable}

\begin{deluxetable}{lccccc}
\tabletypesize{\footnotesize}
\tablecaption{Absolute magnitude of IW\,And-type stars \label{Mv}
			}
\tablecolumns{9}
\tablewidth{0pt}
\tablehead{ 
	\colhead{Name} & \colhead{Dist.(pc)} & \colhead{$A_V$} & \colhead{$m_V$} & \colhead{$M_V$} \\
          }
\startdata
IW And& 835.1 & 0.152\tablenotemark{a} & 14.565\tablenotemark{c} &4.804 \\ 
HO Pup& 618.6 & 0.091\tablenotemark{a} & 13.740\tablenotemark{b} &4.692 \\ 
IM Eri& 191.2 & 0.000\tablenotemark{a} & 11.771\tablenotemark{b} &5.364 \\ 
FY Vul& 603.8 & 1.095\tablenotemark{a} & 13.750\tablenotemark{d} &3.751 \\
V507 Cyg& 613.2 & 0.578\tablenotemark{a} & 14.753\tablenotemark{c} & 5.237 \\
ST Cha& 715.4 & 0.460\tablenotemark{b} & 14.250\tablenotemark{b} &4.517 \\ 
V513 Cas& 845.0 & 2.159\tablenotemark{a} & 15.870\tablenotemark{c} &4.077 \\
KIC 9406652 & 351.5 & 0.030\tablenotemark{a} & 11.710\tablenotemark{b} & 3.905 \\
\enddata
\tablecomments{Distances were converted from Gaia DR2 parallaxes \citep{gai18}.}
\tablenotetext{a}{\citet{gre19}} 
\tablenotetext{b}{\citet{jay18}} 
\tablenotetext{c}{\citet{hen15}}
\tablenotetext{d}{\citet{alf12}}
\end{deluxetable}

\subsection{Photometric Characteristics}

Two major phenomena of HO Pup, i.e., brightenings and deep dips, were found to be consistent with models proposed from two different teams -- \cite{kim20} and \citet{ham14}. The former team explained the usual status of HO Pup (without presenting the deep dips) based on a tilted accretion disk model inspired by the idea proposed by \citet{kat19}. The latter team reproduced both of the sudden brightening and deep dip by variations of the mass transfer rate.

In the tilted accretion disk model, the accretion flow from the donor star will directly fall into the inner disk as well as outer disk.  With a steady supply of mass, the inner part of accretion disk of IW\,And-type stars can stay in a hot state in both quasi-standstill and brightening phases. Once the mass threshold was achieved in the outer disk, brightening could occur from inside to outside of the disk, also known as inside-out brightening \citep{kim20,cou20}.  The brightening event, in fact, shows up in the light curve (for example, see Figure \ref{CFHT} at JD 2458565). Due to the instability of the accretion rate, the entire disk can be cooled since there is little mass left in the disk, and hence the brightness can drop dramatically. 

The performance of the thermal-viscous instability in the tilted accretion disk proposed by \citet{kim20} is consistent with the brightening events and the 3.9-day cycle seen in the light curve of HO\,Pup. Indeed, \citet{kim20} stated the tilted disk models with $10^{17}$~$\mathrm{g s}^{-1}$ were similar to the "heartbeat-type oscillations" of HO\,Pup. Furthermore, in their highly-tilted disk models with accretion rate of $10^{16.75}$ and $10^{17}$~$\mathrm{g s}^{-1}$ (the C1 and C2 model, respectively), the sawtooth-like pattern shown in the quasi-standstill phase resembling the light curve of HO\,Pup displaying the 3.9-day variations (See Figure~\ref{days_scale}). However, as pointed out by the referee, the 3.9-day variations of the HO\,Pup light curve could also be caused by the precession of the tilted disk (the super-orbital modulation). Nevertheless, the tilted-disk models did not include eclipsing-like deep-dip events ($>$ 1\,mag) just after the brightening phase due to the difficulty of reproducing such a one-time dip event in the model. 

To explain the similar phenomenon as seen in the light curve of V513\,Cas, a kind of mass-transfer-burst model has been framed by \citet{ham14}, which reproduces the eclipsing-like deep dip events.  With a sudden change of mass-transfer rate, the deep dip showed up in this model, which is similar to HO\,Pup with drops of brightness by $\sim2.5$~mag. It could be the same mechanism responsible for the observed light curves of either V513\,Cas or HO\,Pup, or both. The event profiles of this model seem to be consistent with observations of an abrupt increase followed by a falling down below the average. Given that the brightness of HO\,Pup dropped by $\sim2.5$~mag during the deep-dip events, this implies a significant change of the mass transfer rate, as the luminosity dropped by 10 times during these events. However, we did not find any evidence that the possible mechanisms of mass-transfer bursts, which are listed by \citet{ham14}, work in HO\,Pup. Also, because of the change of the mass-transfer rate, it is expected the moving average of optical luminosity over a long timescale would be dramatically altered, but no such trend was seen in the HO\,Pup data (even with a very long baseline). Perhaps long-term high-cadence observations could help in resolving the physical reasons behind these deep-dip events (such as detecting the negative superhumps), especially when such observations are carried out during those stochastic events.

Finally, we pointed out that the brightening cycle of HO\,Pup varied from 23 to 61 days, which is neither shown in the models from \citet{kim20} nor \citet{ham14}. If the disk is tilted, the cycle of brightening may change because the geometry of the disk, the tilt angle or the thickness of the disk, could change on a long timescale. Furthermore, while the deep-dip events were observed in 2017-2018 season, the cycle of the brightening before each deep-dip event also has a very large variation, lasting from 10 day to 30 days. A good scenario for explaining such variability with cyclic duration of brightening is yet to be constructed.  

\subsection{Spectroscopic Characteristics}\label{sec:spectralChara}

As shown in lower panel of Figure~\ref{5m}, the spectra of HO\,Pup displayed broad Balmer absorption lines up to H$\zeta$, as well as other helium and metal lines including \ion{He}{1}, \ion{He}{2}, \ion{Fe}{1} and \ion{Mg}{1}. Moreover, we found CV signatures with weak emission cores presented in most of the broad Balmer absorption lines from our P200/DBSP spectrum, which was taken at the brightening right after the maximum brightness. \citet{szk13} observed two CVs, IW\,And  \& V513\,Cas, in their brightening phase, in which their Balmer lines were nearly the same as in HO\,Pup.  Beyond the hydrogen features, there is a double-peaked, bumpy and shoulder-like feature in between H$\beta$ and H$\gamma$, which include the lines of \ion{C}{3}, \ion{N}{3} (known as Bowen fluorescence at 4640~\AA) and \ion{He}{2} (4731~\AA).  This is a unique blended-emission line that can be revealed especially during the phase of CV brightening.  In the phase of quasi-standstill, the Bowen features become much more insignificant as seen in the BOAO/BOES spectrum, along with shallower Balmer absorption. \citet{hes84} reported the time-resolved spectroscopic observations from brightening to quasi-standstill for a classical CV SS\,Cyg, showing the strength of the emissions from Bowen fluorescence declining to the continuum level. The similarity of HO\,Pup spectra features and these CV strongly suggested that HO\,Pup is indeed an IW\,And-type star. Two additional spectroscopic differences between brightenings and quasi-standstill phase not seen in the literature are the weaker \ion{H}{1} 6678 \AA~line shown in the quasi-standstill phase and the \ion{Na}{1} doublet showing up during the brightenings (See Figure~\ref{5m}).

\subsection{Polarimetric Characteristics}

Based on the polarimetric observations taken during the quasi-standstill phase (see upper panel of Figure~\ref{5m}), HO\,Pup exhibited an intrinsic optical polarization variability, at which the polarized light included the light from both the star and the disk. Presence of magnetic field or disk scattering may account for the observed polarization. Despite the variation of polarization level, the polarization percentage $P$ at four different nights were all measured to be less than $1\%$ in our r and i band TRIPOL2 data. This level of polarization has been observed in other three DNe during their quasi-standstill phase \citep{szk82}. 

\subsection{Similarity to Be stars}

Previously, HO\,Pup was considered as a Be star based on its GCAS-like light curves. However, those incomplete light curves were sparsely sampled in the past, as shown in Figure~\ref{allLC}, which could be responsible for the mis classification of HO\,Pup as a Be star. HO\,Pup has better observation coverage since 2016, therefore its DN nature is revealed by \citet{kim20} from the dense sampling of the light curve.

Since both of the Be stars and IW\,And-type stars are hot objects with hydrogen and helium lines, as well as exhibit irregular and abrupt photometric variations (e.g., due to presence of a disk), these two types of stars share a number of common observational features. As evident from the case of HO\,Pup, an IW\,And-type star could be misclassified as Be stars based on limited observational evidence (and/or vice versa). In Table~\ref{OC}, we compared and summarized some of the important observational features of Be stars, IW\,And-type stars and HO\,Pup. For example, Be stars could display regular dip events due to eclipsing as in the case of RW\,Tau with dips fainter by $\sim4$~mag \citep{boo77}. On the other hand, the deepest dip event observed on IW\,And-type stars is fainter by $\sim3$~mag \citep{kat19} without cyclic variations. The observed dip events of HO\,Pup, without regularity of the cycle of the dip events (from 12 to 30 days), resembled higher similarity to IW\,And-type stars rather than eclipsing Be stars. Furthermore, for the brightening events that were not followed by dip events, IW\,And-type stars have much more regularity in their brightening phase (with brightening cycles range from 15 to 200 days) when comparing to Be stars displaying GCAS-type light curve, in which the brightenings are more secular. 

Although the brightening duration of IW\,And-type stars are mostly much shorter than Be stars, they have similar inverse correlation between photometric brightness and spectroscopic line emission.  For IW\,And-type stars, the brightness variation is dominated by accretion disk along with the growth of the disk atmosphere, so we see the growth of the hydrogen absorption.  On the other hand, disk extinction dominates the brightness variation for Be stars \citep{sig13}. After the disk dissipates, we see brighter Be stars with less surrounded amount of line-emitting materials.  The inside-out mass processing is the mechanism a Be star forming a "decretion" disk surrounding the host star \citep{riv13}.  It is very interesting to see the two very different kinds of disks (accretion disk and decretion disk) showing similarly photometric and spectroscopic behavior.

We see a similar interplay in the eight nights of our CFHT/ESPaDOnS observations. During these observations, HO\,Pup experienced a transition from 14 to 13.5~mag at the same time the EW of H$\alpha$ line strength decreased from $-6.4$\AA~to $-1.2$\AA~(see Figure~\ref{CFHT}). A similar behavior between H$\alpha$ emission line strength and photometric brightness was also clearly observed in another Be star, V438\,Aur \citep{lab17}.  However, the obscured phase of HO\,Pup is up to $\sim36$ days which is far less than V438\,Aur (more than 2000 days).

Therefore, as summarized in Table~\ref{OC}, we cannot easily distinguish IW\,And-type stars from the field (luminous) Be stars without knowing the distance, because they share the same color, and nearly the same spectral features.  With the low sampling of light curves and limited spectroscopic observations, neither photometric nor spectroscopic data can be used to distinguish possible IW\,And-type stars from a list of faint Be stars.  If we observe any other unknown IW\,And-type stars with a low-sampling light curve (as in the case of HO\,Pup before 2010), there is a large probability that these stars would be classified as Be stars based on their GCAS-like light curve. Spectroscopic observations might also be difficult to separate these two classes of stars, unless the spectra were taken during the brightening phase which contain unique features of metal lines (such as Bowen fluorescence) for the CV. Since the brightening phases of a CV last only a few days, as compared to Be stars that can last for longer than 10 days, there is a narrow observing window for taking the spectra for these objects.

\begin{deluxetable}{lccc}
\tabletypesize{\footnotesize}
\tablecaption{The spectroscopic and photometric features shown in Be stars, (confirmed) IW\,And-type stars (excluding HO\,Pup) and HO\,Pup  \label{OC}
			}
\tablecolumns{9}
\tablewidth{0pt}
\tablehead{ 
	\colhead{} & \colhead{Be Stars} & \colhead{IW\,And Stars} & \colhead{HO Pup} 
          }
\startdata
Number      & $\sim$3000 & 7+7\tablenotemark{a} & 1 \\ 
Hydrogen lines   & yes & yes & yes \\ 
Helium lines   & yes & yes & yes \\ 
Spectral type & early & early & early \\
Bowen fluorescence & no & yes & yes \\
Dip $\delta M_V$   & 4 mag (RW Tau) & 2.5 mag & 2.5 mag \\ 
Dip cycle   &  $\cdots$    & 15-55 days (ST Cha) & 7-30 days \\
brightening $\delta M_V$  & $<$ 0.5 mag & 0.5-1 mag & 0.5 mag \\ 
brightening cycle & 71 days ($\delta$ Sco) & 15-100 days & 23-61 days \\
brightening duration  & 2-1000 days & a few days & $\sim$5 days 
\enddata
\tablenotetext{a}{The first "7" represents the confirmed IW\,And-type stars as mentioned in the Introduction; the second "7" is counting the candidates listed in footnote 1. } 
\end{deluxetable}

\section{SUMMARY} \label{sec:sum}

Instead of classifying HO Pup as a Be candidate, we confirmed that HO\,Pup is an IW\,And-type star based on the light-curve pattern, spectroscopic characterizations along with the Gaia DR2 distance and SED fitting. As in other IW\,And-type stars, light curves of HO\,Pup display various types of stochastic and quasi-periodic variations such as brightenings, dip events and quasi-standstill phase. To shed light on the stellar physics behind the characteristic light curve of IW\,And-type stars, further well-covered spectroscopic monitoring along with intense photometric observations are highly desirable.

\acknowledgments

This work is partly supported by the Ministry of Science and Technology (Taiwan) under grants of 104-2923-M-008-004-MY5, 107-2119-M-008-014-MY2, 107-2119-M-008-012, 108-2811-M-008-546, and 109-2112-M-155-001. We thank Abert Kong for pointing out the DASCH light curve. We also thank the discussions with Yi Chou, Wen-Ping Chen, Shih-Yun Tang, Michihiro Takami, Chi-Hung Yan, Paula Szkody and Melissa Graham on this work, as well as suggestions from an anonymous referee to improve the manuscript.

Based on observations obtained with the Samuel Oschin Telescope 48-inch and the 60-inch Telescope at the Palomar Observatory as part of the Zwicky Transient Facility project. ZTF is supported by the National Science Foundation under Grant No. AST-1440341 and a collaboration including Caltech, IPAC, the Weizmann Institute for Science, the Oskar Klein Center at Stockholm University, the University of Maryland, the University of Washington, Deutsches Elektronen-Synchrotron and Humboldt University, Los Alamos National Laboratories, the TANGO Consortium of Taiwan, the University of Wisconsin at Milwaukee, and Lawrence Berkeley National Laboratories. Operations are conducted by COO, IPAC, and UW. SED Machine is based upon work supported by the National Science Foundation under Grant No. 1106171

This work was partly supported by the GROWTH (Global Relay of Observatories Watching Transients Happen) project funded by the National Science Foundation under PIRE Grant No 1545949. GROWTH is a collaborative project among California Institute of Technology (USA), University of Maryland College Park (USA), University of Wisconsin Milwaukee (USA), Texas Tech University (USA), San Diego State University (USA), University of Washington (USA), Los Alamos National Laboratory (USA), Tokyo Institute of Technology (Japan), National Central University (Taiwan), Indian Institute of Astrophysics (India), Indian  Institute of Technology Bombay (India), Weizmann Institute of Science (Israel), The Oskar Klein Centre at Stockholm University (Sweden), Humboldt University (Germany), Liverpool John Moores University (UK) and University of Sydney (Australia).
 
This research relied on the SIMBAD and VizieR catalogue access tool and the Aladin plot tool at CDS, Strasbourg (France), and NASA ADS bibliographic services.  This publication makes use of data products from the Wide-field Infrared Survey Explorer, which is a joint project of the University of California, Los Angeles, and the Jet Propulsion Laboratory/California Institute of Technology, funded by the National Aeronautics and Space Administration.  We also made use of data collected at Lulin Observatory, partly supported by MoST grant 108-2112-M-008-001. We sincerely thank the staff and queue observers (Chi-Sheng Lin, Hsiang-Yao Hsaio and Wei-Jie Hou) at the Lulin Observatory for carried out the observations with the SLT telescope. We thank the observer at the P200 Telescope, Marianne Heida, for taking the P200/DBSP spectrum. We also thank staffs of BOAO. We appreciate the achieve and online catalogs provided by ASAS-SN, AAVSO and Pan-STARRS, making this investigation possible for in much longer baseline and/or better time coverage. The DASCH project at Harvard is grateful for partial support from NSF grants AST-0407380, AST-0909073, and AST-1313370.


\begin{thebibliography}{}
\bibitem[Alfonso-Garz{\'o}n et al.(2012)]{alf12} Alfonso-Garz{\'o}n, J., Domingo, A., Mas-Hesse, J.~M., et al.\ 2012, \aap, 548, A79
\bibitem[Bellm \& Sesar(2016)]{bellm2016} Bellm, E.~C., \& Sesar, B.\ 2016, pyraf-dbsp: Reduction pipeline for the Palomar Double Beam Spectrograph, ascl:1602.002
\bibitem[Bellm et al.(2019)]{bel19} Bellm, E.~C., Kulkarni, S.~R., Graham, M.~J., et al.\ 2019, \pasp, 131, 018002
\bibitem[Ben-Ami et al.(2012)]{ben12} Ben-Ami, S., Konidaris, N., Quimby, R., et al.\ 2012, \procspie, 844686
\bibitem[Blagorodnova et al.(2018)]{bla18} Blagorodnova, N., Neill, J.~D., Walters, R., et al.\ 2018, \pasp, 130, 035003
\bibitem[Bookmyer(1977)]{boo77} Bookmyer, B.~B.\ 1977, \pasp, 89, 533
\bibitem[Chambers et al.(2016)]{cha16} Chambers, K.~C., Magnier, E.~A., Metcalfe, N., et al.\ 2016, arXiv e-prints, arXiv:1612.05560
\bibitem[Cenko et al.(2006)]{cen06} Cenko, S.~B., Fox, D.~B., Moon, D.-S., et al.\ 2006, \pasp, 118, 1396
\bibitem[Court et al.(2020)]{cou20} Court, J.~M.~C., Scaringi, S., Littlefield, C., et al.\ 2020, \mnras, 494, 4656
\bibitem[Cutri et al.(2012)]{cut12} Cutri, R.~M., Wright, E.~L., Conrow, T., et al.\ 2012, Explanatory Supplement to the WISE All-Sky Data Release Product
\bibitem[Donati et al.(1997)]{don97} Donati, J.-F., Semel, M., Carter, B.~D., Rees, D.~E., \& Collier Cameron, A.\ 1997, \mnras, 291, 658 
\bibitem[Donati et al.(2007)]{don07} Donati, J.-F., Jardine, M.~M., Gregory, S.~G., et al.\ 2007, \mnras, 380, 1297 
\bibitem[Geier et al.(2013)]{gei13} Geier, S., Heber, U., Edelmann, H., et al.\ 2013, \aap, 557, A122
\bibitem[Geier et al.(2019)]{gei19} Geier, S., Raddi, R., Gentile Fusillo, N.~P., et al.\ 2019, \aap, 621, A38
\bibitem[Gaia Collaboration et al.(2018)]{gai18} Gaia Collaboration, Brown, A.~G.~A., Vallenari, A., et al.\ 2018, \aap, 616, A1
\bibitem[Gies et al.(2013)]{gie13} Gies, D.~R., Guo, Z., Howell, S.~B., et al.\ 2013, \apj, 775, 64
\bibitem[Graham et al.(2019)]{gra19} Graham, M.~J., Kulkarni, S.~R., Bellm, E.~C., et al.\ 2019, \pasp, 131, 078001
\bibitem[Green et al.(2019)]{gre19} Green, G.~M., Schlafly, E., Zucker, C., et al.\ 2019, \apj, 887, 93
\bibitem[Grindlay et al.(2012)]{gri12} Grindlay, J., Tang, S., Los, E., et al.\ 2012, New Horizons in Time Domain Astronomy, 29
\bibitem[Hambsch(2012)]{ham12} Hambsch, F.-J.\ 2012, Journal of the American Association of Variable Star Observers (JAAVSO), 40, 1003
\bibitem[Hameury \& Lasota(2014)]{ham14} Hameury, J.-M., \& Lasota, J.-P.\ 2014, \aap, 569, A48
\bibitem[Henden et al.(2015)]{hen15} Henden, A.~A., Levine, S., Terrell, D., et al.\ 2015, American Astronomical Society Meeting Abstracts \#225
\bibitem[Heinze et al.(2018)]{hei18} Heinze, A.~N., Tonry, J.~L., Denneau, L., et al.\ 2018, \aj, 156, 241
\bibitem[Hessman et al.(1984)]{hes84} Hessman, F.~V., Robinson, E.~L., Nather, R.~E., et al.\ 1984, \apj, 286, 747
\bibitem[Huang(2019)]{hua19} Huang, P., 2019, PhD thesis, Graduate Institute of Astronomy, National Central University
\bibitem[Jayasinghe et al.(2018)]{jay18} Jayasinghe, T., Kochanek, C.~S., Stanek, K.~Z., et al.\ 2018, \mnras, 477, 3145
\bibitem[Kaiser et al.(2010)]{kai10} Kaiser, N., Burgett, W., Chambers, K., et al.\ 2010, \procspie, 77330E
\bibitem[Kato(2019)]{kat19} Kato, T.\ 2019, \pasj, 71, 20
\bibitem[Kim et al.(2002)]{kim02} Kim, K.-M., Jang, B.-H., Han, I., Jang, J. G., Sung, H. C., and Chun, M.-Y. \ 2002, Journal of the Korean Astronomical Society, 35, 221 
\bibitem[Kimura et al.(2020a)]{kim20} Kimura, M., Osaki, Y., Kato, T., et al.\ 2020a, \pasj, 72, 22
\bibitem[Kimura et al.(2020b)]{kim20b} Kimura, M., Osaki, Y., \& Kato, T.\ 2020b, \pasj, 72, 94
\bibitem[Kochanek et al.(2017)]{koc17} Kochanek, C.~S., Shappee, B.~J., Stanek, K.~Z., et al.\ 2017, \pasp, 129, 104502
\bibitem[Koester(2010)]{koe10} Koester, D.\ 2010, \memsai, 81, 921
\bibitem[Kukarkin et al.(1971)]{kuk71} Kukarkin, B.~V., Kholopov, P.~N., Pskovsky, Y.~P., et al.\ 1971, General Catalogue of Variable Stars, 0
\bibitem[Kulkarni (2013)]{kul13} Kulkarni, S.~R. \ 2013, The Astronomer's Telegram, 4807
\bibitem[Kurucz(1993)]{kur93} Kurucz, R.\ 1993, ATLAS9 Stellar Atmosphere Programs and 2 km/s grid. Kurucz CD-ROM No. 13. Cambridge, 13
\bibitem[Labadie-Bartz et al.(2017)]{lab17} Labadie-Bartz, J., Pepper, J., McSwain, M.~V., et al.\ 2017, \aj, 153, 252
\bibitem[Lasker et al.(2008)]{las08} Lasker, B.~M., Lattanzi, M.~G., McLean, B.~J., et al.\ 2008, \aj, 136, 735
\bibitem[Law et al.(2009)]{Law2009} Law, N.~M., Kulkarni, S.~R., Dekany, R.~G., et al.\ 2009, \pasp, 121, 1395
\bibitem[Mainzer et al.(2011)]{mai11} Mainzer, A., Bauer, J., Grav, T., et al.\ 2011, \apj, 731, 53
\bibitem[Manek(1997)]{man97} Manek, J.\ 1997, Information Bulletin on Variable Stars, 4476, 1
\bibitem[Masci et al.(2019)]{mas19} Masci, F.~J., Laher, R.~R., Rusholme, B., et al.\ 2019, \pasp, 131, 018003
\bibitem[Mason \& Howell(2016)]{mas16} Mason, E., \& Howell, S.~B.\ 2016, \aap, 589, A106
\bibitem[Meinunger(1980)]{mei80} Meinunger, L.\ 1980, Information Bulletin on Variable Stars, 1795, 1
\bibitem[Morrissey et al.(2007)]{mor07} Morrissey, P., Conrow, T., Barlow, T.~A., et al.\ 2007, \apjs, 173, 682 
\bibitem[Ngeow et al.(2019)]{nge19} Ngeow, C.-C., Lee, C.-D., Yu, P.-C., et al.\ 2019, Journal of Physics Conference Series, 012010
\bibitem[Ochsenbein et al.(2000)]{och00} Ochsenbein, F., Bauer, P., \& Marcout, J.\ 2000, \aaps, 143, 23
\bibitem[Oke \& Gunn(1982)]{oke82} Oke, J.~B., \& Gunn, J.~E.\ 1982, \pasp, 94, 586
\bibitem[Pojmanski(1997)]{poj97} Pojmanski, G.\ 1997, \actaa, 47, 467
\bibitem[Pojmanski(2002)]{poj02} Pojmanski, G.\ 2002, \actaa, 52, 397
\bibitem[Pojmanski et al.(2005)]{poj05} Pojmanski, G., Pilecki, B., \& Szczygiel, D.\ 2005, \actaa, 55, 275
\bibitem[Rigault et al.(2019)]{rig19} Rigault, M., Neill, J.~D., Blagorodnova, N., et al.\ 2019, \aap, 627, A115
\bibitem[Ritter et al.(2014)]{rit14} Ritter, A., Ngeow, C.~C., Konidaris, N., et al.\ 2014, Contributions of the Astronomical Observatory Skalnate Pleso, 43, 209
\bibitem[Rivinius et al.(2013)]{riv13} Rivinius, T., Carciofi, A.~C., \& Martayan, C.\ 2013, \aapr, 21, 69
\bibitem[Samus et al.(2017)]{sam17} Samus', N.~N., Kazarovets, E.~V., Durlevich, O.~V., Kireeva, N.~N., \& Pastukhova, E.~N.\ 2017, Astronomy Reports, 61, 80 
\bibitem[Sato et al.(2019)]{sat19} Sato, S., Chieh Huang, P., Chen, W.~P., et al.\ 2019, Research in Astronomy and Astrophysics, 19, 136
\bibitem[Schlegel \& Honeycutt(2019)]{sch19} Schlegel, E.~M. \& Honeycutt, R.~K.\ 2019, \apj, 876, 152
\bibitem[Schmidt et al.(1992)]{sch92} Schmidt, G.~D., Elston, R., \& Lupie, O.~L.\ 1992, \aj, 104, 1563
\bibitem[Shappee et al.(2014)]{sha14} Shappee, B.~J., Prieto, J.~L., Grupe, D., et al.\ 2014, \apj, 788, 48
\bibitem[Sigut, \& Patel(2013)]{sig13} Sigut, T.~A.~A., \& Patel, P.\ 2013, \apj, 765, 41
\bibitem[Skrutskie et al.(2006)]{sku06} Skrutskie, M.~F., Cutri, R.~M., Stiening, R., et al.\ 2006, \aj, 131, 1163
\bibitem[Stellingwerf(1978)]{ste78} Stellingwerf, R.~F.\ 1978, \apj, 224, 953
\bibitem[STScI Development Team(2013)]{sts13} STScI Development Team\ 2013, pysynphot: Synthetic photometry software package, ascl:1303.023
\bibitem[Szkody et al.(1982)]{szk82} Szkody, P., Michalsky, J.~J., \& Stokes, G.~M.\ 1982, \pasp, 94, 137
\bibitem[Szkody et al.(2013)]{szk13} Szkody, P., Albright, M., Linnell, A.~P., et al.\ 2013, \pasp, 125, 142
\bibitem[Tremblay \& Bergeron(2009)]{tb09} Tremblay, P.-E. \& Bergeron, P.\ 2009, \apj, 696, 1755
\bibitem[Wolf et al.(2018)]{wol18} Wolf, C., Onken, C.~A., Luvaul, L.~C., et al.\ 2018, \pasa, 35, e010
\bibitem[Yu et al.(2015)]{yu15} Yu, P.~C., Lin, C.~C., Chen, W.~P., et al.\ 2015, \aj, 149, 43 
\bibitem[Yu et al.(2016)]{yu16} Yu, P.-C., Lin, C.-C., Lin, H.-W., et al.\ 2016, \aj, 151, 121 
\bibitem[Yu et al.(2018)]{yu18} Yu, P.-C., Yu, C.-H., Lee, C.-D., et al.\ 2018, \aj, 155, 91 
\bibitem[Zacharias et al.(2005)]{zac05} Zacharias, N., Monet, D.~G., Levine, S.~E., et al.\ 2005, VizieR Online Data Catalog, I/297
\bibitem[Zacharias et al.(2012)]{zac12} Zacharias, N., Finch, C.~T., Girard, T.~M., et al.\ 2012, VizieR Online Data Catalog, I/322A
\end{thebibliography}
\end{document}